\documentclass[11pt,oneside]{article}
\usepackage{setspace,graphicx,amssymb,amsmath,latexsym,amsfonts,amscd,amsthm,multirow,ctable,mathdots,caption,array}
\usepackage{fancyhdr,tabularx,cite,mathrsfs}
\usepackage[headings]{fullpage}
\usepackage{stmaryrd}
\usepackage{rotating}
\usepackage{authblk}
\usepackage{hyperref}
\usepackage{comment,enumitem}

\newcommand{\bea}{\begin{eqnarray}} 
\newcommand{\eea}{\end{eqnarray}} 
\newcommand{\bee}{\begin{eqnarray*}} 
\newcommand{\eee}{\end{eqnarray*}} 
\newcommand{\al}{\begin{align*}} 
\newcommand{\eal}{\end{align*}} 
\newcommand{\be}{\begin{equation}} 
\newcommand{\ee}{\end{equation}} 
\newcommand{\eq}[1]{(\ref{#1})} 
\newcommand{\bem}{\begin{pmatrix}} 
\newcommand{\eem}{\end{pmatrix}}

\def\b{\beta} 
\def\c{\gamma} 
\def\d{\delta} 
 
\def\e{\epsilon}    
\def\f{\phi}  
   
\def\h{\eta} 
\def\i{\iota}

\def\j{\psi} 
\def\k{\kappa}
\def\l{\lambda} 
\def\m{\mu} 
\def\n{\nu} 
 
\def\o{\omega}  
 
\def\p{\pi}    
\def\pa{\partial}        

\def\r{\rho}                  
            
\def\t{\tau} 
\def\th{\theta} 
\def\til{\tilde}

\def\x{\xi} 
 
\def\z{\zeta}

\def\O{\Omega}

\def\Srt{\Phi}

\newcolumntype{R}{ >{$}r <{$}}
\newcolumntype{C}{ >{$}c <{$}}
\newcolumntype{L}{ >{$}l <{$}}
\newcolumntype{F}{>{\centering\arraybackslash}m{1.5cm}}

\def\ll{\ell}

\newcommand{\RR}{{\mathbb R}}
\newcommand{\CC}{{\mathbb C}}
\newcommand{\ZZ}{{\mathbb Z}}
\newcommand{\FF}{{\mathbb F}}

\newcommand{\Aut}{\operatorname{Aut}}

\newcommand{\tr}{\operatorname{{tr}}}

\newcommand{\Sym}{{\textsl{Sym}}}

\newcommand{\SL}{\operatorname{\textsl{SL}}}      

\newcommand{\G}{\Gamma}	
\newcommand{\g}{\gamma}	

\newcommand{\rs}{{X}}	

\newcommand{\Co}{\textsl{Co}}	

\newcommand{\eg}{{\bf EG}}

\newcommand{\B}{\mathbf{b}} 
\newcommand{\C}{\mathbf{c}} 

\newcommand{\PE}{\text{P}}
\newcommand{\AP}{\text{A}}

\newcommand{\FE}{\text{F}}   
\newcommand{\BO}{\text{B}} 
\newcommand{\tot}{\text{tot}}
\newcommand{\gh}{\text{gh}}
\newcommand{\cyl}{\text{cyl}}

\newcommand{\smallcirc}{\kern0.7ex\vcenter{\hbox{\circle{1.5}}}\kern0.2ex}

\newcommand{\normord}[1]{\vcentcolon\mathrel{#1}\vcentcolon}\providecommand{\vcentcolon}{\mathrel{\mathop{:}}} 

\newcommand{\circcolon}{\mathrel{\vcenter{\offinterlineskip\ialign{##\cr$\smallcirc$\cr\noalign{\kern1.5pt}$\smallcirc$\cr}}}} 

\newcommand{\canord}[1]{\circcolon\mathrel{#1}\circcolon}\providecommand{\circcolon}{\mathrel{\mathop{\circcolon}}} 

\theoremstyle{definition}

\theoremstyle{remark}

\numberwithin{equation}{section}

\begin{document}

\setstretch{1.4}

\title{
\vspace{-35pt}
    \textsc{\huge{ $K3$ Elliptic Genus and \\an Umbral Moonshine Module
    }  }
}

\author[1]{Vassilis Anagiannis\thanks{v.anagiannis@uva.nl}}
\author[1,2]{Miranda C. N. Cheng\thanks{mcheng@uva.nl}}
\author[3,4]{Sarah M. Harrison\thanks{sarharr@physics.mcgill.ca}}

\affil[1]{Institute of Physics, University of Amsterdam, Amsterdam, the Netherlands}
\affil[2]{Korteweg-de Vries Institute for Mathematics, University of Amsterdam, Amsterdam, the Netherlands}
\affil[3]{Center for the Fundamental Laws of Nature\\ 
Harvard University, Cambridge, MA 02138, USA}
\affil[4]{Department of Mathematics and Statistics and Department of Physics, McGill University, Montreal, QC, Canada}
\date{}

\maketitle

\abstract{
Umbral moonshine connects the symmetry groups of the 23 Niemeier lattices with 23 sets of distinguished mock modular forms. 
The 23 cases of umbral moonshine have a uniform relation to symmetries of $K3$ string theories. 
Moreover, a supersymmetric vertex operator algebra with Conway sporadic symmetry also enjoys a close relation to the $K3$ elliptic genus.
 Inspired by the above two relations between moonshine and $K3$ string theory, we construct a chiral CFT by orbifolding the free theory of 24 chiral fermions and two pairs of fermionic and bosonic ghosts. In this paper we mainly focus on the case of umbral moonshine corresponding to the Niemeier lattice with root system given by 6 copies of $D_4$ root system. 
 This CFT then leads  to the  construction of an infinite-dimensional graded module for the umbral group $G^{D_4^{\oplus 6}}$ whose graded characters coincide with the umbral moonshine functions. We also comment on how one can recover all umbral moonshine functions corresponding to the Niemeier root systems $A_5^{\oplus 4}D_4$, $A_7^{\oplus 2}D_5^{\oplus 2}$ , $A_{11}D_7 E_6$, $A_{17}E_7$, and $D_{10}E_7^{\oplus 2}$. 
}

\clearpage

\tableofcontents

\clearpage

\section{Introduction}
\label{sec:intro}
The moonshine phenomenon, the study of which began with the discovery of monstrous moonshine \cite{CN}, describes an interesting connection between modular objects and finite groups. 
Recent years have seen a surge of interest in moonshine, initiated by the observation \cite{EOT} that the elliptic genus of $K3$ surfaces
 has a  close relation to the representation theory of the sporadic finite group $M_{24}$. 
It was soon realised that this $M_{24}$ connection is but one of the 23 instances of the umbral moonshine \cite{UM,UMNL}, which associates distinguished mock modular forms to elements of finite groups arising from the symmetries of specific lattices.

Recall that there are 24 inequivalent positive-definite, even unimodular lattices with rank 24. Among them, the Leech lattice $\Lambda_{\text{Leech}}$ is distinguished by the property that it has no root vectors. The other 23, which we call the Niemeier lattices $N^X$, are uniquely labelled by their root systems $X$. 
For each of the 23 $N^X$, we are interested in the finite group  $G^X:=\Aut(N^X)/{\rm Weyl}(X)$. For each element $g\in G^X$, umbral moonshine then attaches a vector-valued mock modular form $H^X_g=(H^X_{g,r})$,  $r\in \ZZ/2m$ where $m$ denotes the Coxeter number of $X$, such that it coincides with the graded character of an infinite-dimensional module of $G^X$ which we refer to as the umbral moonshine module. 
More explicitly, the main conjecture of umbral moonshine states that there exists an naturally-defined $\ZZ$-graded $G^X$-module 
$$K^X_r=\bigoplus_{\substack{D\in\ZZ_{\geq   0} \\ D=-r^2\rm{mod}~ 4m}} K^X_{r,D}$$ for $r\in\{1,\dots,m\}$ such that $H^X_g=(H^X_{g,r})$ for $1\leq r\leq m$ is given by 
\be\label{eqn:umbralmodule} 
H^X_{g,r}(\t) = -2\, q^{-{1\over 4m}}\, \d_{r^2,1(4m)} + \sum_{\substack{D\in\ZZ_{\geq  0} \\ D=-r^2\rm{mod} ~4m}} q^{{D\over 4m}}  {\rm{Tr}}_{K^X_{r,D}} (g) ~
\ee
where $q:=e^{2\pi i \tau}$.
 The other components  are then determined by the property that, when combined with the index $m$ theta functions  \eq{theta_m}, the 2-variable functions 
 \be\Psi^X_{g}(\t,\z) := \sum_{r\in \ZZ/2m} H^X_{g,r}(\t) \theta_{m,r}(\t,\z)\ee 
 transform under $\SL_2(\ZZ)$ in a way such that they form examples of the so-called mock Jacobi foms \cite{Dabholkar:2012nd}. Note that the singular term when $\tau \to i\infty$, given by $\mp 2  q^{-{1\over 4m}}$ if $r^2=1\,(4m)$ and vanishes otherwise, is of lowest possible order that is allowed by mock modularity, which means that  $H^X_g$ have Fourier coefficients that grow as slowly as possible given the mock modular properties and are examples of  so-called optimal mock Jacobi forms. This mathematical property has led to the (unique) characterisation and classification of the umbral moonshine mock modular forms  $H^X_g$ \cite{optimal,weight_one_jacobi}. 

The existence of the umbral moonshine modules $K^X=\bigoplus_{1\leq r\leq m} K^X_r$ has been established in \cite{Gannon:2012ck,Duncan:2015rga} for all 23 $X$, though their construction is still unknown in general.  
To construct and to understand these modules is arguably the most important open problem in the study of umbral moonshine. To the best of our knowledge, a uniform construction of all 23 umbral moonshine modules, which is clearly desirable given the uniform construction of the moonshine relation itself, is not yet in sight. 
In the case of classical monstrous moonshine as well as in Conway moonshine, the underlying modules famously possess the structure of a vertex operator algebra (or, more loosely speaking, a chiral CFT) \cite{FLM,conway}. 
In the umbral case, despite the proof that $K^X$ has to share certain crucial features of a vertex operator algebra \cite{GMM,GUM}, its precise structure is not yet clear. 
That said, encouraging results have been obtained for certain cases of umbral moonshine with fairly small groups\footnote{With no more than 24 elements, they are small compared to $|M_{24}|\sim 10^8$.}: the cases where the Niemeier root systems $X$ are given by $E_8^{\oplus 3}$ \cite{Duncan:2014tya}, $A_6^{\oplus 4}$ and  $A_{12}^{\oplus 2}$ \cite{Duncan:2017bhh}, as well as $D_6^{\oplus 4}$, $D_8^{\oplus 3}$, $D_{12}^{\oplus 2}$ and $D_{24}$ \cite{Dtypemodule}.  
The construction in \cite{Duncan:2014tya} is rather different from those in \cite{Duncan:2017bhh,Dtypemodule}: the former relies on special identities satisfied by the mock modular forms $H^{E_8^{\oplus 3}}_{g,r}$  and the latter constructs $H^{X}_{g,r}$ through the mock Jacobi forms $\Psi^X_{g}$ and their associated meromorphic Jacobi forms. 

In this work we take a different route and exploit the relation between the (twined) $K3$ elliptic genus and umbral and Conway moonshine which we now explain.
 As mentioned before, 
the first case of umbral moonshine, the Mathieu moonshine corresponding to $X=A_1^{\oplus 24}$, was discovered in the context of the $K3$ elliptic genus. 
In \cite{UMK3} it was proposed that all 23 cases of umbral moonshine, not just Mathieu moonshine, are relevant for describing the symmetries of $K3$ sigma models. In particular, one can associate in a uniform way, using (twined) elliptic genera of ADE singularities as one of the main ingredients, a weak Jacobi form (of weight 0 and index 1) $\f^X_g$ to each of the 23 root systems $X$ and each conjugacy class $[g]\in G^X$.
This proposal was further tested in \cite{LG} and refined in \cite{k3_lattices}. 

Another moonshine connection to $K3$ sigma model comes from Conway moonshine.  
In \cite{GHV}, a generalisation of the Mukai theorem states that the physical symmetries relevant for twining the $K3$ elliptic genus are given by 4-plane preserving subgroups of the Conway group $\Co_0$, the automorphism group of the Leech lattice $\Lambda_{\text{Leech}}$. 
(Throughout the paper we say that a subgroup of $\Co_0$ or $G^X$ is $n$-plane preserving if it fixes pointwise an $n$-dimensional subspace in the natural $24$-dimensional representation, given by the corresponding lattice $\Lambda_{\text{Leech}}$ or $N^X$.) This classification inspired the interesting construction that associates to each $4$-plane preserving conjugacy class $[g]\in \Co_0$ two (possibly coinciding) weak Jacobi forms \cite{derived}, denoted $\f_{\pm,g}$. Furthermore,  it was proposed that they play the role of twined elliptic genera of $K3$ sigma models. 
Inspired by the above results and relying on various empirical evidence, in \cite{k3_lattices}  (Conjecture 6)  it was conjectured that all the weak Jacobi forms arising from (the 4-plane preserving part of) Conway and umbral moonshine are realised as $K3$ elliptic genera twined by a supersymmetry-preserving symmetry of the sigma model at certain points in the moduli space. Conversely, every $K3$ twined elliptic genus (at any point in  the moduli space) coincides with one of the moonshine Jacobi forms alluded to above. 
Physical arguments given in \cite{Paquette:2017gmb} 
promote this conjecture to a near-theorem.

In this work we mainly focus on the construction of the umbral moonshine module for the case $X=D_4^{\oplus 6}$, with $G^{D_4^{\oplus 6}}\cong 3.S_6$.  The first element of this construction is the following relation between the  the weak Jacobi form $\f^X_g$ arising from umbral moonshine and Conway moonshine. In many (but not all) cases, $\f^X_g$ 
  coincides with one of the functions $\f_{\epsilon,g'}$ arising from Conway moonshine. See Appendix D of \cite{k3_lattices}. In particular, for all $g\in G^{D_4^{\oplus 6}}$ there is some Conway element $g'$ and a certain sign $\epsilon$ such that  $\f^{D_4^{\oplus 6}}_g=\f_{\epsilon,g'}$. It is however important to note that $G^{D_4^{\oplus 6}}\cong 3.S_6$ is a subgroup of $\Co_0$ which is {\em not} 4-plane preserving. The second ingredient is  the construction of a chiral conformal field theory, by taking an $\ZZ/2$-orbifold of 24 free chiral fermions and 2 pairs of fermionic and bosonic ghost fields.
 When graded by the charges of  the ghost $U(1)$ current in a specific way, its twined partition function coincides with $\f_{\epsilon,g}$. The crucial difference, however, is that the symmetry of this chiral theory accommodates the full $\Co_0$ without the 4-plane preserving constraints. As a result, in some sense this chiral CFT ${\cal T}$ plays the role of a bridge between Conway and umbral moonshine. 
 The above two elements together with a special property (cf. \S\ref{subsec:mock_forms} and Conjecture 6.2 of \cite{UMNL}) of the $D_4^{\oplus 6}$ module then leads to a construction of the umbral module. 
We will also comment on how one can recover all $H_g^X$ for $X=$$A_5^{\oplus 4}D_4$, $A_7^{\oplus 2}D_5^{\oplus 2}$ , $A_{11}D_7 E_6$, $A_{17}E_7$, and $D_{10}E_7^{\oplus 2}$, and some of the $H_g^X$ for $X=A_1^{\oplus 24}, A_{2}^{\oplus 12}$, $A_3^{\oplus 8}$, $A_8^{\oplus 3}$, $E_4^{\oplus 6}$ using the same ingredients.

The rest of the paper is organised as follows. 
 In \S\ref{sec:MoonshineK3EG} we review how umbral and Conway moonshine lead to the weak Jacobi forms $\f^X_g$ and $\f_{\pm,g}$ respectively, for every $g\in G^X$ in the former case and every 4-plane preserving element $g$ of $\Co_0$ in the latter case. 
In \S\ref{sec:The Chiral CFT} we present the construction of the chiral conformal field theory ${\cal T}$ and demonstrate that its graded twined partition functions coincide with $\f_{\pm,g}$ when making a specific choice of chemical potentials for the ghost $U(1)$ currents. In \S\ref{sec:module} we combine these ingredients and explicitly describe the $G^{D_4^{\oplus 6}}\text{--action}$ on the infinite dimensional $\ZZ/2$-graded vector space underlying the $D_4^{\oplus 6}$ case of umbral moonshine. In \S\ref{sec:discussions} we describe how to recover umbral moonshine functions for certain other cases of umbral moonshine from the twined parition functions of ${\cal T}$ and the singularity CFTs, and comment on a few open questions. 
Finally we collect details of  the ghost theory and relevant data in the appendices. 

\section{Moonshine and the $K3$ Elliptic Genus}
\label{sec:MoonshineK3EG}
In this section we review the construction developed in \cite{UMK3} and \cite{derived} of certain weak Jacobi forms, playing the role of  twined elliptic genera of $K3$ sigma models \cite{k3_lattices,Paquette:2017gmb}, from umbral and Conway moonshine.

\subsection{Umbral Twining Genera}
\label{subsec:umbral_twining_genera}

For each case of umbral moonshine corresponding to the Niemeier lattice $N^X$ there is a natural way to attach a weak Jacobi form of weight zero and index one to each $g\in G^X$  \cite{UMK3}. 
This weak Jacobi form $\f^X_g$ constitutes two parts: one is the (twined) elliptic genus of the SCFT describing the surface singularities corresponding to the root system $X$, and the second comes from the contribution of the umbral moonshine mock modular forms.

We start by reviewing the first part, the singularity elliptic genus. 
The type of singularities a $K3$ surface can develop are given by the so-called du Val or Kleinein singularities, which admit an ADE classification. These singularities are isomorphic to $\mathbb C^2/\Gamma$, where $\Gamma$ is a finite subgroup of $SU(2)$ as in the McKay correspondence. 
Let $m$ denote the Coxeter number of the corresponding root system. 
Recall that $\mathcal N=2$ superconformal minimal models (which have spectral flow symmetries) also admit an ADE classification \cite{Cappelli:1987xt}, and the central charge $c:=3\hat c$ is given by \be 
 \hat{c}= 1-{2\over m}. 
\ee
The classification of these supersymmetric minimal  models stems from the classification of modular invariant combinations of left- and right-moving characters of affine ${\mathfrak sl}_2$, given in terms of 
a $2m\times 2m$ matrix, which we denote by $\Omega^\Phi$ for the minimal model corresponding to the simply-laced root system $\Phi$. The explicit expression for $\Omega^\Phi$ can be found in \cite{Cappelli:1987xt}.  
In terms of these matrices, the elliptic genus of the super minimal model is given by \cite{witten:LG}
\be
\label{def:EG_minimal_coset}
Z_{\text{minimal}}^\Srt (\tau,\zeta) =  \sum_{r,r' \in \mathbb Z/2m \mathbb Z} \O^\Srt_{r,r'} \til \chi^r_{r'}(\tau,\zeta) ={\text{Tr}( \O^\Srt \cdot \til \chi)} ,
\ee
where $\til \chi^r_s (\tau,\zeta)$, with $|s| \leq r-1 < m$, are corresponding minimal model characters.

In \cite{OV}, a 2d CFT description of type II string theory compactified on $\mathbb C^2/\Gamma$ was proposed to
be given by an $\ZZ_m$--orbifold of the corresponding supersymmetric minimal model tensored with a non-compact CFT, and 
 takes the form
\be\label{eq:OV}
\left( {\cal N}=2~\text{minimal } \otimes  {\cal N}=2~ \left (\frac{\SL(2,\RR)}{U(1)}\right)_m~ {\text{coset}} \right) /{\ZZ_m},
\ee
where the second factor,
  the $\left (\frac{\SL(2,\RR)}{U(1)}\right )_m$ supercoset model,
   describes the geometry of a semi-infinite cigar  \cite{Witten:1991yr} and has central charge
 \[
\hat{c} = 1+ \frac{2}{m}.
\] 
The spectrum of the theory contains a discrete part as well as a continuous part; the latter exists due to the fact that the theory is non-compact and gives a non-holomorphic (in the $\tau$-variable) contribution to the elliptic genus of the theory \cite{Troost:2010ud}. 
Both mathematically and physically, there is a well-defined way to isolate the holomorphic part of the elliptic genus, which we denote by $Z_{L_m}$, corresponding to the contribution from the discrete part of the spectrum. 
It is given by \cite{Troost:2010ud,Eguchi:2010cb,Ashok:2011cy}
\be
Z_{L_m}(\tau,\zeta)=\frac{1}{2} \sum_{s=1}^{m}{\rm Ch}^{(\til R)}_{\text{massless}} (\tau,\zeta;m+2-s)+{\rm Ch}^{(\til R)}_{\text{massless}} (\tau,\zeta;s)= \frac{1}{2}  \m_{m,0}\big(\t,\frac{\z}{m}\big) \frac{i\theta_1(\tau,\zeta)}{\eta(\tau)^3}.
\ee
In the above equation, we make use of the Ramond character graded by $(-1)^F$ 
\[
{\rm Ch}^{(\til R)}_{\rm massless}(\tau,\zeta;s) =\frac{i \th_1(\tau,\zeta)}{\eta^3(\t)} \sum_{k\in \ZZ}y^{2k} q^{mk^2}\frac{(yq^{mk})^{\frac{s-1}{m}}}{1-yq^{mk}}
\]
where $s/2$ is the $U(1)$ charge of the highest weight,
and the (specialized) Appell--Lerch sum 
\be\label{def:mu_0}
\m_{m,0}(\tau,\zeta) = - \sum_{k\in\mathbb Z} q^{m{k}^2}y^{2km}\frac{1+yq^k }{1-yq^k}  
\ee
which is responsible for the mock modularity of $Z_{L_m}$.

Putting it together using the ``orbifoldization formula"\cite{KYY} , the (holomorphic part of the) elliptic genus of the orbifold theory is given by 
\begin{align}\label{def:EG_ADE}
{\bf EG}(\tau,\zeta;\Phi) &=  \frac{1}{m} \sum_{a,b \in \ZZ/m\ZZ} q^{a^2} y^{2a} \, Z_{\rm minimal}^\Srt(\tau,\zeta+a\t+b) Z_{L_m}(\tau,\zeta+a\t+b) . 
\end{align}
See also \cite{Murthy:2013mya,Ashok:2013pya,Harvey:2014nha}.
In this paper we  use the following definition: given $X=\Phi_1\oplus \Phi_2 \oplus \dots$  a union of simply-laced root systems $\Phi_i$ with the same Coxeter number, we write ${\bf EG}(X) := {\bf EG}(\Phi_1) + {\bf EG}(\Phi_2) + \dots$. For instance, when $X= {D_4}^{\oplus 6}$, we have ${\bf EG}(\tau,\zeta;{D_4}^{\oplus 6}) = 6 \,{\bf EG}(\tau,\zeta;{D_4})$.

Now we discuss the second part of the weak Jacobi form $\phi^X_g$, arising from the contribution of the umbral moonshine mock modular forms. 
It is shown in \cite{UMK3} that for each of the 23 Niemeier lattices $N^X$, the following function
\be
\phi^X_e(\t,\z) :=  {\bf EG}(\tau,\zeta;X) + {\theta_1^2(\tau,\zeta) \over 2\,\eta^6(\tau)} \left({1\over 2\pi i}{\partial\over \partial \omega} \Psi^X_e(\tau,\omega)\right) \Big\lvert_{\omega=0},
\ee
is always equal to ${\bf EG}(\t,\z;K3)$. In other words, the above expression gives us 23 ways to split ${\bf EG}(K3)$ into a part given by singularity elliptic genus and a part given by umbral moonshine mock modular forms. Moreover, for each $g\in G^X$ umbral moonshine gives us a natural analogue for the second part by replacing $\Psi^X_e$ with the graded character $\Psi^X_g$ of the umbral moonshine module. At the same time, the explicit $G^X$--action on the Niemeier root system $X$ translates into an  $G^X$--action on the singularity CFT which preserves its superconformal structure,  and leads to a definition of its twined elliptic genus ${\bf EG}_g(X) := \text{Tr}(g \dots)$. 
As a result, it is natural to define
\be\label{eq:um_twining}
\phi_g^X(\tau,\zeta) := \eg_g(\t,\z;X) + {\theta_1^2(\tau,\zeta) \over 2\,\eta^6(\tau)} \left({1\over 2\pi i}{\partial\over \partial w} \Psi^X_g(\tau,w)\right) \Big\lvert_{w=0}.
\ee 
It can be shown that for all 23 Niemeier root systems $X$ and $g\in G^X$ the above definition leads to a weak Jacobi form for certain $\Gamma_g\subseteq \SL_2(\ZZ)$, possibly with a non-trivial multiplier system. As mentioned in \S\ref{sec:intro}, when $g$ is a 4-plane preserving element, these play the role of twined elliptic genera that are realised at certain points in the moduli space of $K3$ sigma models.

\subsection{Conway Twining Genera}
\label{subsec:conway}

In \cite{conway} Duncan and Mack-Crane constructed a $\frac{1}{2}\ZZ$-graded infinite-dimensional $\Co_0$-module $V^{s\natural}$, given by the quantum states of a chiral CFT. 
The starting point is a theory of 24 free real
chiral fermions, which we then orbifold by a $\ZZ_2$ which acts with a minus sign on all 24
fermionic fields. 
The Neveu-Schwarz and Ramond sectors of the original theory before orbifolding split into $\AP=\AP^0\oplus\AP^1$ and $\PE=\PE^0\oplus\PE^1$ respectively, where the superscripts $0,1$ denote the invariant and anti-invariant subspaces under the orbifold action and $\AP$, $\PE$ denote the anti-periodic and periodic boundary conditions of the fermions on the \emph{cylinder}. 
Define \begin{equation}
\label{eq:conway_modules}
V^{s\natural}:=\AP^0\oplus\PE^1,~~~V^{s\natural}_{\text{tw}}:=\AP^1\oplus\PE^0.
\end{equation}
It was then shown that $V^{s\natural}$ (the NS sector) has the structure of a super vertex operator algebra and $V^{s\natural}_{\text{tw}}$  (the Ramond sector) is a canonically twisted $V^{s\natural}$-module. 
Moreover, both $V^{s\natural}$ and $V^{s\natural}_{\text{tw}}$ admit a 
$\Co_0$ action, as can be seen from identifying the 24-(complex) dimensional space in the description of the VOA of 24 free fermions with $\Lambda_{\text{Leech}} \otimes_{\ZZ} \CC$. 
Conway moonshine then states that the partition functions of $V^{s\natural}$ and $V^{s\natural}_{\text{tw}}$ twined by elements of $\Co_0$ are (up to a scaling $\t\mapsto 2\t$) normalised Hauptmoduls of genus zero subgroups of $\SL_2(\RR)$. We refer to \cite{conway} for further details.

Given a fixed $n$-dimensional subspace in $\Lambda_{\text{Leech}} \otimes_{\ZZ} \RR$, there are different ways to build $U(1)$ currents from the fermions of  $V^{s\natural}$\cite{m5,derived}. 
Here we are interested in the case when a $U(1)$ current $J$ is constructed from fermions associated to a subspace of dimension $n=4$. Fixing this $U(1)$  together with the compatibility with ${\cal N}=1$ supersymmetry breaks the symmetry of the theory from $\Co_0$ to the subgroup of $\Co_0$ that preserves the given 4-plane. Conversely, given a 4-plane preserving $G\subset \Co_0$  one can construct a $U(1)$ current $J$ such that $V^{s\natural}_{\text{tw}}$,  when equipped with a module structure for  $J$ and for the ${\cal N}=1$ superconformal algebra, has symmetry $G$. 
Interestingly, the $U(1)$-charged partition function of $V^{s\natural}_{\text{tw}}$ coincides with ${\bf EG}(K3)$ (up to a sign) \cite{derived}. 
  More generally, one can consider the  $U(1)$-graded character of the twisted Conway module
twined by a 4-plane preserving element of $\Co_0$: 
\begin{equation}
\f_g:=-\tr_{V^{s\natural}_{\text{tw}}}\left[\mathfrak{z}~\hat{g}~y^{J_0}q^{L_0-\frac{c}{24}}\right],
\end{equation}
where $J_0$ is the zero mode of the aforementioned $U(1)$ current. 
In the above $\hat{g}$ denotes the \emph{lift} of $g$ from $\text{SO}(24)$ to $\text{Spin}(24)$, which is necessitated by the fact that the ground states in the Ramond sector form a $4096$-dimensional spinor representation (Clifford module) that we denote by {\bf CM}, and $\mathfrak{z}$ is the lift of $-\text{Id}\in\Co_0$.
Explicitly, it is given by
\begin{equation}
\label{eq:phi_g}
\begin{split}
\f_g(\tau,\zeta)&=\frac{1}{2}\left[\frac{\theta_3(\tau,\zeta)^2}{\theta_3(\tau,0)^2}\frac{\eta_{-g}(\tau/2)}{\eta_{-g}(\tau)}-\frac{\theta_4(\tau,\zeta)^2}{\theta_4(\tau,0)^2}\frac{\eta_{g}(\tau/2)}{\eta_{g}(\tau)}\right. \\
&~~~~~~~~~~\left.-\frac{\theta_2(\tau,\zeta)^2}{\theta_2(\tau,0)^2}C_{-g}\eta_{-g}(\tau)-\frac{\theta_1(\tau,\zeta)^2}{\eta(\tau)^6}D_g\eta_g(\tau)\right].
\end{split}
\end{equation}

The notation used above is as follows.  For each $g\in\Co_0$ denote by $\lambda^{\pm1}_i$ with $i=1,\dots,12$, the pairs of complex conjugate pairs of eigenvalues of the unique non-trivial 24-dimensional representation, arising from the natural action of $\Co_0\cong\text{Aut}(\Lambda_{\text{Leech}})$ on the Leech lattice. Then define the quantities
\begin{align}
\label{eq:definitions_eta_C_D}
\begin{split}
&\h_{\pm g}(\t):= q\prod_{n=1}^{\infty}\prod_{i=1}^{12}\left(1\mp\l_iq^n\right)\left(1\mp\l_i^{-1}q^n\right), \\
&C_{-g}:=\n\prod_{i=1}^{12}\left(1+\l_i^{-1}\right)=\prod_{i=1}^{12}\left(\n_i+\n_i^{-1}\right)~.
\end{split}
\end{align}
where we have set $\n_i=\l_i^{1/2}$ and $\n=\prod_{i=1}^{12}\n_i$. Also note that $C_{-g}=\tr_{\text{\bf CM}}\hat{g}$ and this is what determines the branch choice of $\nu_i$. See \cite{derived} for more details. Now assume that $g\in\Co_0$ fixes at least a $4$-plane in the $24$-dimensional representation, and define
\begin{equation}
D_g:=\n\sideset{}{'}\prod_{i=1}^{12}\left(1-\l_i^{-1}\right)=\sideset{}{'}\prod_{i=1}^{12}\left(\n_i-\n_i^{-1}\right).
\end{equation}
where $\sideset{}{'}\prod$ skips two pairs of eigenvalues for which $\l_i^{\pm1}=1$. Notice that $D_g$ is non-vanishing if and only if it fixes exactly a $4$-plane. In the latter case, $D_g$ is determined up to a sign by the eigenvalues of $g$, since we are free to exchange what we call $\l_i$ and $\l_i^{-1}$. As a result, for exactly four-plane preserving elements there are in fact two choices of $\f_g$ depending on the choice of the sign of $D_{g}$, and we define
\begin{equation}
\begin{split}\label{eq:phi_epsilon}
\phi_{\e,g}(\tau,\zeta)&:=\frac{1}{2}\left[\frac{\theta_3(\tau,\zeta)^2}{\theta_3(\tau,0)^2}\frac{\eta_{-g}(\tau/2)}{\eta_{-g}(\tau)}-\frac{\theta_4(\tau,\zeta)^2}{\theta_4(\tau,0)^2}\frac{\eta_{g}(\tau/2)}{\eta_{g}(\tau)}\right. \\
&~~~~~~~~~~\left.-\frac{\theta_2(\tau,\zeta)^2}{\theta_2(\tau,0)^2}C_{-g}\eta_{-g}(\tau)-\e~\frac{\theta_1(\tau,\zeta)^2}{\eta(\tau)^6}\left|D_g\right|\eta_g(\tau)\right].
\end{split}
\end{equation}
where $\e=\pm1$ encodes the sign ambiguity of $D_g$.
In all cases, it was shown that 
$\f_g$ are Jacobi forms of weight $0$ and index $1$, at some level, for every $g\in\Co_0$ that fixes at least a 4-plane. As discussed in \S\ref{sec:intro}, they play the role of a twined $K3$ elliptic genus in the study of symmetries of $K3$ sigma models.

\section{The Chiral CFT}
\label{sec:The Chiral CFT}

In this section we present the construction of a chiral CFT ${\cal T}$, by $\ZZ_2$-orbifolding a free theory consisting of 12 complex chiral fermions, 2 fermionic and 2 bosonic ghost systems.
Its symmetries accommodate the umbral group we are interested in, and its twined partition functions reproduce (among others) the weak Jacobi forms $\f_g$ reviewed in \S\ref{sec:MoonshineK3EG}.

\subsection{The Fermions}
\label{subsec:the_fermions}
The first ingredient to build our chiral theory ${\cal T}$ is 24 real chiral fermions $\til{\psi}_1,\dots,\til{\psi}_{24}$, similar to the starting point of the Conway module discussed in \S\ref{subsec:conway}. 
Equivalently, this theory, which we call $\mathcal{T}_\j$, is given by 12 complex chiral fermions, $\psi_{i}^\pm :={1\over \sqrt{2}} (\til\psi_i \pm i \til\psi_{i+12})$,  with the action
\begin{align}
S_\j=\frac{1}{4\p}\int d^2z\sum_{i=1}^{12}\left(\j^+_i\bar{\pa}\j^-_i+\j^-_i\bar{\pa}\j^+_i\right).
\end{align}
Their OPEs take the form
\begin{align}
\j^\pm_i(z)\j^\pm_j(z^\prime)\sim\mathcal{O}(z-z^\prime),~~~\j^\pm_i(z)\j^\mp_j(z^\prime)\sim\frac{\d_{ij}}{z-z^\prime}.
\end{align}
The associated Viraroso operator is given by
\begin{align}
L^\j=\sum_{n\in\ZZ}L^\j_nz^{-n-2}=-\frac{1}{2}\canord{\left(\j^+_i\pa\j^-_i+\j^-_i\pa\j^+_i\right)},
\end{align}
with respect to which $\j^\pm_i$ are holomorphic primary fields with weight $1/2$. 
The open dots denote the regular part of the associated expression; we refer to this as the canonical ordering. In terms of modes, it means that the annihilators are always put to the right.
By expanding the fields in modes,
\begin{align}
\j^\pm_i(z)=\sum_{r}\j^\pm_{i,r}~z^{-r-\frac{1}{2}},
\end{align}
the OPEs lead to the standard anti-commutation relations
\begin{align}
\left\lbrace\j^\pm_{i,r}\j^\pm_{j,s}\right\rbrace=0~,~~~\left\lbrace\j^\pm_{i,r}\j^\mp_{j,s}\right\rbrace=\d_{ij}\d_{r+s,0}.
\end{align}
The  ${\SL}(2,\RR)$-invariant vacuum $|0\rangle$ satisfies the usual highest weight condition
\begin{align}
\j^\pm_{i,r}|0\rangle=0~~~\forall~r>0.
\end{align}
Note that here the canonical ordering coincides with the  normal ordering corresponding to the above canonical vacuum, where the positive modes are annihilators.

To compute the characters of the theory,
consider the conformal mapping from the complex plane to the cylinder given by $z=e^w$. 
We denote the Virasoro zero mode on the cylinder by 
\begin{align}
L^\j_{\cyl,0}:=L^\j_0-\frac{c_\j}{24} 
\end{align}
where $c_\j=12$ is the central charge. 
Consider general boundary conditions parametrized by $\r$
\begin{align}
\label{eq:general_bc_fermions}
\j^\pm_i(w+2\p i)=e^{\mp2\p i\r}\j^\pm_i(w). 
\end{align}
The periodic ($\PE$, $\r=0$) and anti-periodic ($\AP$, $\r=1/2$)  cases correspond to the usual Ramond and NS sectors. Note that the $\psi^\pm_i$ must acquire opposite phases as in \eqref{eq:general_bc_fermions} so that the Virasoro operator remains periodic. 

The $\AP$ sector Hilbert space $\mathcal{T}_{\j,\AP}$ is built by acting on the ground state $|0\rangle$ with the creation operators $\psi^{\pm}_{i,r}$ with $r\leq -1/2$, and its character is given by  
\begin{align}
\chi_\j^\AP(\t):=\tr_{\mathcal{T}_{\j,\AP}}\left[q^{L^\j_{\cyl,0}}\right]=q^{-1/2}\prod_{n=1}^\infty\prod_{i=1}^{12}(1+q^{n-1/2})^2=\left(\frac{\th_3(\t,0)}{\h(\t)}\right)^{12}.
\end{align}
We also define an operator $(-1)^F$ which has the property that it anticommutes with all the fermionic modes, squares to the identity, and acts trivially on $|0\rangle$. 
Note that $(-1)^F$ commutes with the Virasoro operator and hence we can define the following character,
\begin{align}
\til\chi_\j^\AP(\t):=\tr_{\mathcal{T}_{\j,\AP}}\left[(-1)^F q^{L^\j_{\cyl,0}}\right]=q^{-1/2}\prod_{n=1}^\infty\prod_{i=1}^{12}(1-q^{n-1/2})^2=\left(\frac{\th_4(\t,0)}{\h(\t)}\right)^{12}.
\end{align} 

In the $\PE$ sector $\mathcal{T}_{\j,\PE}$, the ground states form a  $2^{12}$-dimensional representation of the $24$-dimensional Clifford algebra. Explicitly, a basis can be given by  the mononomials
\begin{align}
\label{eq:mononomials}
\psi^-_{i_1,0}\cdots\psi^-_{i_k,0}|s\rangle,
\end{align}
where we single out $|s\rangle$ to be annihilated by all the $\psi^+_{i,0}$ and we require that $(-1)^F$ acts trivially on $|s\rangle$. 
The conformal weight of the $\PE$ ground states is equal to $\frac{24}{16}=\frac{3}{2}$, as each of the complex fermions (along with its complex conjugate) contributes $\frac{2}{16}$ due to the presence of twist fields that interpolate between the $\AP$ and $\PE$ sectors.
Putting things together, we obtain the following $\PE$ sector characters. 
\begin{align}
\chi_\j^\PE(\t):=\tr_{\mathcal{T}_{\j,\PE}}\left[q^{L^\j_{\cyl,0}}\right]=2^{12}q\prod_{n=1}^\infty\prod_{i=1}^{12}(1+q^n)^2=\left(\frac{\th_2(\t,0)}{\h(\t)}\right)^{12},
\end{align}
\begin{align}
\til\chi_\j^\PE(\t):=\tr_\PE\left[(-1)^F q^{L^\j_{\cyl,0}}\right]
=\left(\frac{\th_1(\t,0)}{\h(\t)}\right)^{12}=0.
\end{align}
The latter vanishes because half of the ground states have $-1$ eigenvalue under $(-1)^F$, while the rest have $+1$.

Later we will consider an orbifold of ${\cal T}_\psi$ by a $\ZZ_2$ generated by $\x$, which acts on the fermions by
\begin{align}
\x\j^\pm_i=-\j^\pm_i,
\end{align}
and trivially on the $\AP$ ground state.
Note that any state with an odd (resp. even) number of excitations is an eigenstate of $\x$ with eigenvalue $-1$ (resp. $+1$),  and hence $\x$ acts in exactly the same way as $(-1)^F$ on the quantum states in both $\AP$ and $\PE$ sectors. 
Therefore we conclude that
\begin{align}
\tr_{\mathcal{T}_{\j,\AP}}\left[\x~q^{L^\j_{\cyl,0}}\right]=\til\chi_\j^\AP(\t)=\left(\frac{\th_4(\t,0)}{\h(\t)}\right)^{12},
\end{align}
\begin{align}
\tr_{\mathcal{T}_{\j,\PE}}\left[\x~q^{L^\j_{\cyl,0}}\right]=\til\chi_\j^\PE(\t)=0.
\end{align}

\subsection{The Ghosts}
\label{subsec:ghosts}
The next ingredients we  need are the fermionic and bosonic ghost systems (see \cite{friedan}, \cite{Kausch,Kausch2,ghosts,ghost_vertex,ghost_bosonic} and references therein for related discussions). 
They are described by the action
\begin{align}
S_\gh=\frac{1}{2\pi}\int d^2z~\B\bar{\pa}\C,
\end{align}
where $\B$ and $\C$ are holomorphic fields of weights $h$ and $1-h$ respectively. We  focus on the cases where $h\in\frac{1}{2}\ZZ$. Since there are many similarities between the fermionic and bosonic cases, we  use the boldface notation to refer to either. When we need to make the distinction, we  use $b$, $c$ to denote the fermionic ghosts and $\b$, $\g$ to denote the bosonic ghosts. We will also use a parameter $\k$, which equals $+1$ for the fermionic case and $-1$ for the bosonic case.

The OPEs between the ghost fields have the form
\begin{align}
\B(z)\C(w)\sim\frac{\k}{z-w},~~~\B(z)\B(w)=\C(z)\C(w)\sim\mathcal{O}(1)
\end{align}
and the Virasoro operator is given by
\begin{equation}
\label{eq:virasoro_ghosts}
L^\gh=(1-h)\canord{(\pa\B)\C}-h\canord{\B(\pa\C)},
\end{equation}
with respect to which $\B$, $\C$ are primary. The central charge of the ghost system is then given by
\begin{align}
\label{eq:ghost_c}
c_{\B\C}=\k(1-3Q^2),
\end{align}
where we have introduced $Q:=\k(1-2h)$ for later convenience.
The mode expansions on the complex plane are
\begin{align}\label{eq:modee_ghosts}
\B(z)=\sum_r\B_rz^{-r-h}~,~~~\C(z)=\sum_r\C_rz^{-r-(1-h)},
\end{align}
and canonical quantization leads to the (anti)commutation relations
\begin{align}
\left\lbrace\B_r,\C_s\right\rbrace_\k:=\B_r\C_s+\k\C_s\B_r=\k\d_{r+s,0}.
\end{align}
From (\ref{eq:virasoro_ghosts}) and (\ref{eq:modee_ghosts}) we see that 
 the $SL(2,\RR)$-invariant vacuum $|0\rangle$ is determined by the highest weight condition
\begin{align}
\begin{split}
\B_r|0\rangle&=0~\forall~r\geq1-h, \\
\C_r|0\rangle&=0~\forall~r\geq h.
\end{split}
\end{align}
Consequently, the canonical ordering does not generally coincide with the usual normal ordering. 
As before, we consider the $\r$-twisted sectors for the ghost systems  
corresponding to the boundary conditions
\begin{align}
\label{eq:general_bc_ghosts}
\B(w+2\pi i)=e^{-2\pi i\r}\B(w),~~~\C(w+2\pi i)=e^{2\pi i\r}\C(w),
\end{align}
where $w$ is the natural coordinate on the cylinder and is given by $z=e^w$. The periodic case $\r=0$ corresponds to the $\PE$ sector, while the anti-periodic case $\r=1/2$ corresponds to the $\AP$ sector\footnote{ We introduce both sectors irrespective of the statistics of the fields, since they will both appear when we consider
the $\ZZ_2$ orbifold.}.

The natural ground states on the cylinder are defined as the states annihilated by all positive modes, ${\bf b}_r, \C_r$ with $r>0$.
Note that, for the ghost systems, these are in general different from the $\SL(2,\RR)$-invariant ground state $|0\rangle$. 
The corresponding energies, namely the eigenvalues of $L^\gh_{\cyl,0}=L^\gh_0-c_{\B\C}/24$,
of the $\PE$ and $\AP$ sector ground states are given  by $\frac{\k}{12}$ and $-\frac{\k}{24}$ respectively, as is calculated in appendix \ref{ap:ghost_ground_states}.

Another important feature of the ghost systems is that they have the following  $U(1)$ current 
\begin{align}
J=-\canord{\B\C}.
\end{align}
In fact, as we show in appendix \ref{ap:ghost_ground_states}, the $\AP$ sector ground states for both fermonic and bosonic ghosts are unique. The $\PE$ sector has two degenerate ground states for the fermonic ghost system due to the presence of the fermonic zero modes $b_0, c_0$, while it has a single ground state for the bosonic system. We denote the (unique) $\AP$ ground states for the fermonic (F) and bosonic (B) ghosts by $|\Omega_\AP^{\FE}\rangle$ and $|\Omega_\AP^{\BO}\rangle$, respectively. The (unique) $\PE$ ground state for the bosonic ghost is denoted by $|\Omega_\PE^{\BO}\rangle$, and the two degenerate $\PE$ ground states for the fermionic ghost system are denoted by $|\Omega_{\PE,\pm}^{\FE}\rangle$. They are distinguished by
\begin{align}
\begin{split}
b_0|\Omega_{\PE,-}^{\FE}\rangle=0,~~~b_0|\Omega_{\PE,+}^{\FE}\rangle=|\Omega_{\PE,-}^{\FE}\rangle,~~~c_0|\Omega_{\PE,+}^{\FE}\rangle=0,~~~c_0|\Omega_{\PE,-}^{\FE}\rangle=|\Omega_{\PE,+}^{\FE}\rangle.\\
\end{split}
\end{align}

Next we derive the characters of the ghost systems, defined by 
\begin{align}
\chi_a^\text{S}(\tau,\zeta):=\tr_\text{S}\left[y^{J_{\cyl,0}}~q^{L^\gh_{\cyl,0}}\right],
\end{align}
where $\text{S}=\lbrace\PE,\AP\rbrace$ denotes the sector and $a=\lbrace\FE,\BO\rbrace$ distinguishes between the fermionic and bosonic ghosts, respectively.
Note that the other commonly used character, defined by $\tr_\text{S}\left[(-1)^{J_{\cyl,0}}y^{J_{cyl,0}}~q^{L^\gh_{\cyl,0}}\right]$, is simply given by the above by a shift $\z\mapsto \z+{1\over2}$.

Building on the unique ground state $|\Omega_\AP^{F}\rangle$, the $\AP$ sector Hilbert space of the fermionic ghost system leads to the character 
\begin{align}
\chi_\FE^\AP(\tau,\zeta)=q^{-1/24}\prod_{n=1}^\infty\left(1+yq^{n-1/2}\right)\left(1+y^{-1}q^{n-1/2}\right)=\frac{\th_3(\tau,\zeta)}{\h(\t)}.
\end{align}
Similarly, by accounting for all possible states in the Fock space created by the negative integral modes of the ghost fields $b$, $c$ acting on both of the ground states $|\Omega_{\PE,\pm}^{F}\rangle$, we obtain the character
\begin{align}
\chi_\FE^\PE(\tau,\zeta)=q^{1/12}\left(y^{1/2}+y^{-1/2}\right)\prod_{n=1}^\infty\left(1+yq^n\right)\left(1+y^{-1}q^n\right)=\frac{\th_2(\tau,\zeta)}{\h(\t)}. 
\end{align}

For the bosonic ghost system, the $\AP$ sector character is given similarly by 
\begin{align}
\chi_\BO^\AP(\tau,\zeta)=q^{1/24}\prod_{n=1}^\infty\left(1-yq^{n-1/2}\right)^{-1}\left(1-y^{-1}q^{n-1/2}\right)^{-1}=\frac{\h(\t)}{\th_4(\tau,\zeta)},
\end{align}
In the $\PE$ sector, care has to be taken due to the presence of the bosonic zero mode $\gamma_0$. As we will see in \S\ref{subsec:twinings} (also see \cite{ghosts}), the contribution of $\gamma_0$ can be regularised and the total character is  given  by 
\begin{align}
\label{eq:chi_B_R_z}
\chi_\BO^\PE(\tau,\zeta)=q^{-1/12}y^{1/2}(1-y)^{-1}\prod_{n=1}^\infty\left(1-yq^n\right)^{-1}\left(1-y^{-1}q^n\right)^{-1}=i\frac{\h(\t)}{\th_1(\tau,\zeta)},
\end{align}

Finally, we would like to consider a $\ZZ_2$-orbifold of the ghost systems, where the non-trivial group action is given by $\x\B=-\B$ and $\x\C=-\C$. 
The resulting characters will be related to the characters $\chi_a^\text{S}(\tau,\zeta+1/2)$ we calculated above, since the action of the corresponding group element $\x$ corresponds to including the operator $(-1)^{J_{\cyl,0}}$, similarly to the case of the  chiral fermions discussed in \S\ref{subsec:the_fermions}. The only nontrivial part of this implementation is the sign of the ground state(s) under $\x$, which is analysed in appendix \ref{ap:ghost_orbifold}. 
 The results are given by
\begin{align}
\til\chi^{\text{S}}_\FE(\tau,\zeta):=\tr_{\text{S}}\left[\x~y^{J_{\cyl,0}}q^{L^\gh_{\cyl,0}}\right]=(-1)^{h-\frac{1}{2}}~\chi_\FE^{\text{S}}(\tau,\zeta+1/2)
\end{align}
for the fermionic ghosts, while for the bosonic ghosts we have
\begin{align}
\til\chi^{\text{S}}_\BO(\tau,\zeta):=\tr_{\text{S}}\left[\x~y^{J_{\cyl,0}}q^{L^\gh_{\cyl,0}}\right]=(-1)^{3h+\frac{1}{2}}~\chi_\BO^{\text{S}}(\tau,\zeta+1/2),
\end{align}
where $\text{S}$ denotes either of the two sectors. Notice that all the characters we have computed in this section do not depend on the central charge of the ghost systems. However, we will see that by requiring the final CFT to have certain supersymmetry we can completely fix the central charge of the ghosts systems.

\subsection{The Orbifold Theory}
\label{subsec:orbifold/susy}
After describing the basic ingredients, we  now put them together and construct the chiral CFT that will reproduce the K3 elliptic genus and its twinings. Let $\cal{T}_\BO$ denote the theory of 2 copies of the bosonic ghost system and the theory of 2 copies of the fermionic ghost system. We will later see that ${\cal T}$ can be equipped with an $\mathcal{N}=4$ superconformal symmetry for  certain choices of $h_\BO$ and $h_\FE$. 
We will let  $h_\BO=\frac{1}{2}$ and $h_\FE=1$, corresponding to the total central charge $c_\mathcal{T}=-4-2+12=6$.

We want to consider a $\ZZ_2$ orbifold of the theory
\begin{align}
\mathcal{T}^{\text{free}}=\mathcal{T}_\BO\otimes\mathcal{T}_\FE\otimes\mathcal{T}_\j,
\end{align}
where $\ZZ_2=\lbrace1,\x\rbrace$ acts on the individual components of $\mathcal{T}^{\text{free}}$ as we have described in the previous sections. Specifically, we want to consider
\begin{equation}
\mathcal{T}=\left(\mathcal{T}_{\BO,\AP}^0\otimes\mathcal{T}_{\FE,\AP}^0\otimes\mathcal{T}_{\j,\AP}^0\right)\oplus\left(\mathcal{T}_{\BO,\PE}^1\otimes\mathcal{T}_{\FE,\PE}^1\otimes\mathcal{T}_{\j,\PE}^1\right).
\end{equation}
where the $0,1$ superscripts denote respectively the invariant and anti-invariant part under the orbifold, in the corresponding sector denoted by the subscript. Notice that $\mathcal{T}_{\j,\AP}^0\oplus\mathcal{T}_{\j,\PE}^1$ is isomorphic (as a VOA) to the Conway module $V^{s\natural}$ \eqref{eq:conway_modules}. 

Introducing chemical potentials $y_1=e^{2\pi i\z_1}$ and $y_2=e^{2\pi i\z_2}$ for the bosonic and fermionic ghosts respectively, we now define the following partition function
\begin{align}
\label{eq:partition_function_total_theory}
Z(\tau,\z_1,\z_2):=\tr_{\mathcal{T}}\left[y_1^{J_{\cyl,0}^\BO}y_2^{J_{\cyl,0}^\FE}q^{L_0^\tot-\frac{6}{24}}\right],
\end{align}
where $L_0^\tot$ is the total Virasoro zero mode of the theory, and $J_{\cyl,0}^\BO=J_{\cyl,0}^{\BO,1}+J_{\cyl,0}^{\BO,2}$, and $J_{\cyl,0}^\FE=J_{\cyl,0}^{\FE,1}+J_{\cyl,0}^{\FE,2}$ are the zero modes of the $U(1)$ currents of the two bosonic and two fermionic ghosts, respectively. Using the results of the previous sections, we compute
\begin{equation}
\begin{split}
&Z(\tau,\zeta_1,\z_2)=\tr_{\mathcal{T}_\AP^{\text{free}}}\left[\frac{1}{2}(1+\x)y_1^{J_{\cyl,0}^\BO}y_2^{J_{\cyl,0}^\FE}q^{L_0^\tot-\frac{6}{24}}\right]+\tr_{\mathcal{T}_\PE^{\text{free}}}\left[\frac{1}{2}(1-\x)y_1^{J_{\cyl,0}^\BO}y_2^{J_{\cyl,0}^\FE}q^{L_0^\tot-\frac{6}{24}}\right] \\
&=\frac{1}{2}\left[\frac{\th_3(\tau,\zeta_2)^2}{\th_4(\tau,\zeta_1)^2}\left(\frac{\th_3(\tau,0)}{\h(\t)}\right)^{12}-\frac{\th_4(\tau,\zeta_2)^2}{\th_3(\tau,\zeta_1)^2}\left(\frac{\th_4(\tau,0)}{\h(\t)}\right)^{12}-\frac{\th_2(\tau,\zeta_2)^2}{\th_1(\tau,\zeta_1)^2}\left(\frac{\th_2(\tau,0)}{\h(\t)}\right)^{12}\right].
\end{split}
\end{equation}
We observe that, by specializing to $\z_1=1/2$ and $\z_2:=\z$, we retrieve the K3 elliptic genus in the non-standard form presented in \cite{derived}
\begin{align}
\label{eq:elliptic_Z}
\begin{split}
&\textbf{EG}(\tau,\zeta;K3)=Z\left(\t,\frac{1}{2},\z\right)=\tr_{\mathcal{T}}\left[(-1)^{J_{\cyl,0}^\BO}~y^{J_{\cyl,0}^\FE}~q^{L_0^\tot-\frac{6}{24}}\right] \\
&=\frac{1}{2}\left[\frac{\th_3(\tau,\zeta)^2}{\th_3(\t,0)^2}\left(\frac{\th_3(\tau,0)}{\h(\t)}\right)^{12}-\frac{\th_4(\tau,\zeta)^2}{\th_4(\t,0)^2}\left(\frac{\th_4(\tau,0)}{\h(\t)}\right)^{12}-\frac{\th_2(\tau,\zeta)^2}{\th_2(\t,0)^2}\left(\frac{\th_2(\tau,0)}{\h(\t)}\right)^{12}\right].
\end{split}
\end{align}

In the remainder of this section we  discuss the supersymmetry of $\mathcal{T}$. We will see that an  $\mathcal{N}=4$ superconformal  structure is  possible when the central charge of the total theory is $c_{\cal T}=6$. In the following sections we will not make use of this   $\mathcal{N}=4$ structure. 
First, one can equip  the theory $\mathcal{T}_\j$ of $12$ complex fermions with the structure of $\mathcal{N}=1$ superconformal field theory \cite{superconway}. To preserve this $\mathcal{N}=1$ superconformal symmetry, the manifest $\text{Spin(24)}$ symmetry is broken to $\Co_0$. Moreover, it is also possible equip  $\mathcal{T}_\j$ with an $\mathcal{N}=4$ structure, which breaks  $\Co_0$ symmetry to its 3-plane preserving subgroups depending on the choice of the $SU(2)$ current \cite{m5}. 
One can also equip the whole theory ${\cal T}$, for certain choices of the ghost conformal weights, with an $\mathcal{N}=4$  superconformal symmetry by combining the $\mathcal{N}=4$ structure of $\mathcal{T}_\j$ with an $\mathcal{N}=4$ structure of the ghost theory. In particular, for our choice $h_\BO=1/2$ and $h_\FE=1$ the total theory has an $\mathcal{N}=4$ superconformal algebra at $c=6$, precisely the superconformal symmetry of $K3$ non-linear sigma models.

If the conformal weights of a pairs of $bc-\beta\gamma$ ghost system satisfy $h_\FE=h_\BO+\tfrac{1}{2}$,  there exists an $\mathcal{N}=1$ current of weight $3/2$. In our case with two pairs of $bc-\beta\gamma$ ghosts, it is given by
\begin{equation}
\label{eq:ghost_supercurrent}
G=\sum_{j=1}^2\left(-\frac{1}{2}\partial\b_j c_j+\frac{2h_\FE-1}{2}\partial(\b_jc_j)-2b_j\c_j\right).
\end{equation}
To enhance this to $\mathcal{N}=4$, we need an $SU(2)$ subalgebra generated by $J_i$ with $i=1,2,3$. One can show that such currents are given by 
\begin{equation}
J_1=\frac{i}{2}(\b_1\g_2-\b_2\g_1)~,~~~J_2=\frac{1}{2}(\b_1\g_2+\b_2\g_1)~,~~~J_3=\frac{1}{2}\canord{\b_1\g_1-\b_2\g_2}.
\end{equation}
From now on we will choose 
\begin{equation}
h_\FE=1~,~~h_\BO=\frac{1}{2},
\end{equation}
so that $SU(2)$ cuurent algebra is given by
\begin{equation}
J_i(z)J_j(w)\sim\frac{\d_{ij}~k_\gh/2}{(z-w)^2}+\frac{i\e_{ijk}J_k(w)}{z-w}.
\end{equation}
with $k_\gh=-1$. 
Acting with the generators $J_i$ on $G$ (with $h_\FE=1$) to construct the rest of the supercurrents, we get
\begin{equation}
\begin{split}
J_i(z)G(w)\sim-\frac{i}{2}\frac{1}{z-w}G_i(w).
\end{split}
\end{equation}
One can check that, together with the Virasoro field 
\begin{equation}
L^\gh=\sum_{i=1}^2\canord{-b_i\partial c_i+\frac{1}{2}\partial \b_i \c_i-\frac{1}{2}\b_i\partial\c_i},
\end{equation}
the fields $G$, $G_i$ and $J_i$ indeed form an $\mathcal{N}=4$ SCA with central charge $c_\gh=-6$ and level $k_\gh=-1$. As in \cite{m5} we can further define
\begin{equation}
G_1^\pm:=\frac{1}{\sqrt{2}}(G\pm iG_3),~~G_2^\pm:=\pm\frac{i}{\sqrt{2}}(G_1\pm iG_2),
\end{equation}
which transform in the representation $\mathbf{2}+\bar{\mathbf{2}}$ of $SU(2)$, and reproduce the standard small $\mathcal{N}=4$ SCA.

We note that the supercurrents, as well as the $J_i$, survive the orbifold, since they are all bilinears in the ghost fields. 
In the remaining part of the paper we will however not make use of this superconformal symmetry. In particular, we have used the grading by the ghost $U(1)$-currents instead of the $R$-current in defining  the (twined) partition function \eqref{eq:partition_function_total_theory} and in \S\ref{subsec:twinings}.
Especially, different from the construction in \cite{derived}, where the grading is with respect to a $U(1)$ current built from a choice of four real fermions, the symmetry groups  of our theory ${\cal T}$ are not
restricted to 
be 4-plane preserving subgroups of $\Co_0$. 
 The choice of the central charge of the ghost systems is in any case immaterial for the construction of our module, since the relevant characters do not depend on it.

\subsection{The Twined Characters}
\label{subsec:twinings}
As discussed in the previous subsection, $\mathcal{T}$ has a manifest $\text{Spin}(24)$ symmetry. We can thus consider the twined partition function 
\begin{align}\label{eqn:def_twinedpf}
Z_g\left(\tau,\zeta_1,\z_2\right):=\tr_{\mathcal{T}}\left[g~y_1^{J_{\cyl,0}^\BO}y_2^{J_{\cyl,0}^\FE}q^{L_0^\tot-\frac{6}{24}}\right],
\end{align}
by an element $g\in\text{Spin}(24)$, which has a manifest action on $\mathcal{T}_\j$ (cf. \S\ref{subsec:conway}) and acts trivially on the ghost systems.
We now specialize to the case $g\in\Co_0<\text{Spin}(24)$, i.e. to elements that preserve the $\mathcal{N}=1$ superconformal structure of $\mathcal{T}_\j$.  Specifically, the relevant characters are twined as follows
\begin{align}
\tr_{\mathcal{T}_\AP^{\text{free}}}\left[g~y_1^{J_{\cyl,0}^\BO}y_2^{J_{\cyl,0}^\FE}q^{L_0^\tot-\frac{6}{24}}\right]=\frac{\h_{-g}(\t/2)}{\h_{-g}(\t)}\frac{\th_3(\tau,\zeta_2)^2}{\th_3\left(\tau,\zeta_1-\frac{1}{2}\right)^2},
\end{align}
\begin{align}
\tr_{\mathcal{T}_\AP^{\text{free}}}\left[\x~g~y_1^{J_{\cyl,0}^\BO}y_2^{J_{\cyl,0}^\FE}q^{L_0^\tot-\frac{6}{24}}\right]=-\frac{\h_{g}(\t/2)}{\h_{g}(\t)}\frac{\th_4(\tau,\zeta_2)^2}{\th_4\left(\tau,\zeta_1-\frac{1}{2}\right)^2},
\end{align}
\begin{align}
\tr_{\mathcal{T}_\PE^{\text{free}}}\left[g~y_1^{J_{\cyl,0}^\BO}y_2^{J_{\cyl,0}^\FE}q^{L_0^\tot-\frac{6}{24}}\right]=-C_{-g}\h_{-g}(\t)\frac{\th_2(\tau,\zeta_2)^2}{\th_2\left(\tau,\zeta_1-\frac{1}{2}\right)^2},
\end{align}
\begin{align}
\label{eq:problematic}
\tr_{\mathcal{T}_\PE^{\text{free}}}\left[\x~g~y_1^{J_{\cyl,0}^\BO}y_2^{J_{\cyl,0}^\FE}q^{L_0^\tot-\frac{6}{24}}\right]=q\n\prod_{n=1}^\infty\prod_{i=1}^{12}\left(1-\l_iq^n\right)\left(1-\l_i^{-1}q^{n-1}\right)\frac{\th_1(\tau,\zeta_2)^2}{\th_1\left(\tau,\zeta_1-\frac{1}{2}\right)^2}.
\end{align}
where the factors $\th_i(\tau,\zeta_2)^2/\th_i(\tau,\zeta_1-1/2)^2$ originate from the ghosts contribution.

In order to make contact with K3 and the umbral module discussed in the next section, we further specialize to a subgroup $G$ of  $\Co_0$, such that each $g\in G$ generates a 4-plane preserving subgroup of $\Co_0$. 
Note that by requiring that $g\in G$ is $4$-plane preserving does not imply in general that $G$ is  $4$-plane preserving. For instance, in the case $G\cong 3.S_6$ that is of special interest for us, different $g\in 3.S_6$ do not in general fix the same $4$-plane, and thus $3.S_6$ does not preserve a $4$-plane.

Finally, we specialise the fugacities of the ghost currents to the values
 $\z_2=\z$ and $\z_1=1/2$.
Note that care has to be taken when taking the  $\z_1\to1/2$ limit in \eqref{eq:problematic}.
On the one hand, the degeneracy of A ground states in ${\cal T}_\psi$ and the fact that at least two of the twelve pairs of $g$-eigenvalues are given by unity leads to a zero in the numerator. On the other hand, the infinite degeneracy of bosonic ghost ground states requires regularisation when taking  $\z_1\to1/2$. As a result, we regularise the partition function by introducing an adiabatic shift in boundary condition given by a small positive parameter $\h$. We consider the boundary conditions $\r=0+\h$ and $\r=1/2+\h$ as in \eqref{eq:general_bc_fermions}, \eqref{eq:general_bc_ghosts}, and compute the $\h\rightarrow0^+$ limit of the partition function $Z^\h_g\left(\t,\frac{1}{2},\z\right)$ with the regulator $\h$ present. This is straightforward for all the terms besides \eqref{eq:problematic}, which receives the following contributions
\begin{align}
\begin{split}
&\tilde{\chi}_\BO^{\PE,\h}(\t,1/2)^2=q^{-2/12}(1-q^\h)^{-2}\prod_{n=1}^\infty\left(1-q^{n+\h}\right)^{-2}\left(1-q^{n-\h}\right)^{-2}, \\
&\tilde{\chi}_\FE^{\PE,\h}(\tau,\zeta)^2=-q^{2/12}\left(iy^{1/2}q^\d-iy^{-1/2}\right)^2\prod_{n=1}^\infty\left(1-yq^{n+\h}\right)^2\left(1-y^{-1}q^{n-\h}\right)^2, \\
&\til\chi_\j^{\PE,\h}(\t)=q\n\prod_{n=1}^\infty\left(1-q^{n-\h}\right)^2\left(1-q^{n-1+\h}\right)^2\prod_{i=1}^{10}\left(1-\l_iq^{n-\h}\right)\left(1-\l_i^{-1}q^{n-1+\h}\right).
\end{split}
\end{align}
We see that, upon multiplying the above expressions, the potentially problematic factors $(1-q^\h)^{\pm2}$ 
drop out and we get
\begin{align}
\begin{split}
\lim_{\h\rightarrow0^+}\tr_{\mathcal{T}_\PE^{\text{free}}}\left[\x~g~y_1^{J_{\cyl,0}^\BO}y_2^{J_{\cyl,0}^\FE}q^{L_0^\tot-\frac{6}{24}}\right]&=\lim_{\h\rightarrow0^+}\tilde{\chi}_\BO^{\PE,\h}(\t,1/2)^2~\tilde{\chi}_\FE^{\PE,\h}(\tau,\zeta)^2~\til\chi_\j^{\PE,\h}(\t) \\
&=\frac{\theta_1(\tau,\zeta)^2}{\eta(\tau)^6}D_g\eta_g.
\end{split}
\end{align}
Putting everything together, we finally get
\begin{equation}
\label{eq:final_twining}
\begin{split}
\lim_{\h\to 0^+}Z_g^{\h} \left(\t,\frac{1}{2},\z\right)&=\frac{1}{2}\left[\frac{\theta_3(\tau,\zeta)^2}{\theta_3(\tau,0)^2}\frac{\eta_{-g}(\tau/2)}{\eta_{-g}(\tau)}-\frac{\theta_4(\tau,\zeta)^2}{\theta_4(\tau,0)^2}\frac{\eta_{g}(\tau/2)}{\eta_{g}(\tau)}\right. \\
&~~~~~~~~~~\left.-\frac{\theta_2(\tau,\zeta)^2}{\theta_2(\tau,0)^2}C_{-g}\eta_{-g}(\tau)-\frac{\theta_1(\tau,\zeta)^2}{\eta(\tau)^6}D_g\eta_g(\tau)\right].
\end{split}
\end{equation}

We observe that this agrees with $\f_{\e,g}$ as defined in \eqref{eq:phi_g} and \eqref{eq:phi_epsilon}, the Conway twining graded by a $k=1$ $U(1)$ current. In particular, note that the same sign ambiguity in $D_g$ in the twining of the Conway CFT described in \S\ref{subsec:conway} is also present here, leading to the sign $\epsilon$ in the definition of the twining functions. 
As mentioned before, a crucial difference is that the $U(1)$ grading in the ${\cal T}$ is preserved by the $G$-action since it is constructed out of the ghost fields which the group acts trivially on.

\section{The Module for $D_4^{\oplus 6}$ Umbral Moonshine}
\label{sec:module}

The goal of the section is to explain how the ingredients in the previous sections lead to a $\ZZ_2$-graded infinite dimensional vector space admitting a  $G^{D_4^{\oplus 6}}$--action  that underlies the $D_4^{\oplus 6}$ case of umbral moonshine. In particular, we will describe 
 how the umbral mock modular forms $H^{D_4^{\oplus 6}}_g$ for all elements $g$ of the umbral group $G^{D_4^{\oplus 6}}$ are recovered from the twined partition functions of the chiral CFT ${\cal T}$. In \S\ref{subsec:thegroup} we describe an explicit construction of the group. In \S\ref{sec:Singularities} we explain the action of the group on the BPS states of 6 copies of  the CFT describing a singularity of $D_4$ type. 
In \S\ref{subsec:mock_forms} we combine the ingredients and give expressions for $H^{D_4^{\oplus 6}}_g$  in terms of these physical ingredients.

\subsection{The Group}
\label{subsec:thegroup}
For completeness, in this subsection we describe a concrete realization of the group $3.S_6$, following \cite{wilson}. 
The hexacode is the unique three-dimensional code of length 6 over $\FF_4$  that is Hermitian and self-dual.
It is the glue code of the Niemeier lattice $N^{D_4^{\oplus 6}}$ with root system $D_4^{\oplus 6}$ \cite{sphere_packing}, and for this reason it plays a significant role in the case of umbral moonshine corresponding to  $N^{D_4^{\oplus 6}}$.   
Moreover, this code also plays an important role in the construction of the largest Mathieu group $M_{24}$. 
Its automorphism is given by $3.A_6$, which can be explicitly constructed in the following way. 
Write $\FF_4=\{0,1,\o,\bar \o\}$ with 
$$\o^2 =\bar \o, ~~\bar \o^2 = \o,~~ \o^3= 1. $$
The triple cover of the alternating group $A_6$ can be generated by the permutations $(1,2)(3,4)$, $(1,2)(5,6)$, $(3,4)(5,6)$, $(1,3,5)(2,4,6)$, $(1,3)(2,4)$, as well as the composition of the permutation and multiplication $(1,2,3){\rm diag}(1,1,1,1,\bar \o,\o)$. This group acts on the 6 coordinates and in particular induces all even permutations. It also contains the element corresponding to scalar multiplication by $\o$ and by $\bar \o$. Hence we have constructed a group with centre $3 \cong \ZZ/3$ and we will call $z$ the generator of the center corresponding to ${\rm diag}(\o,\o,\o,\o,\o,\o)$, the scalar multiplication by $\o$. 

The group $3.A_6$ can be enlarged to $3.S_6$ by adjoining an extra generator which acts on a vector in $\FF_4^6$ by permuting the last two coordinates followed by a complex conjugation: $\o \leftrightarrow \bar\o$. This group is often referred to as the semi-automorphism group of the hexacode, since it leaves the code invariant but does not act linearly on it. For this reason, the group $G^{D_4^{\oplus 6}} := {\rm Aut}(N^{D_4^{\oplus 6}})/({\rm Weyl}({D_4}))^{\otimes 6} \cong 3.S_6 $ is the  umbral group corresponding to the corresponding case of umbral moonshine.

From the above description, we can define a representation for the group $G^{D_4^{\oplus 6}}$ given by the group homomorphism $\epsilon: G^{D_4^{\oplus 6}} \to \{1,-1\}$, given by   $\epsilon_g = 1$ (-1) when $g$ induces an even (odd) permutation on the 6 coordinates. In the notation in Table \ref{tab:chars:irr:6+3}, this is given by the irreducible character $\chi_2$. 
This representation will play an important role in describing the umbral module in what follows. 

More generally, the action of $G^{D_4^{\oplus 6}}$ on $\FF_4^6$ determines the umbral moonshine module for the ${D_4^{\oplus 6}}$ case of umbral moonshine. 
For later use we will now describe this action in more detail. Writing the natural basis of $\FF_4^6$ as given by ${\bf e}_0^i$, ${\bf e}_1^i$, ${\bf e}_\o^i$ and ${\bf e}_{\bar \o}^i$ for $i=1,\dots,6$, we obtain a 24-dimensional permutation representation of $G^{D_4^{\oplus 6}}$. The corresponding 24-dimensional cycle shape is denoted by $\widetilde{\Pi}_g$ in Table \ref{tab:chars:eul:6+3}. Furthermore, from the above construction of $G^{D_4^{\oplus 6}}$ it is clear that the action of $G^{D_4^{\oplus 6}}$ does not mix ${\bf e}_0^i$ with  ${\bf e}_1^i$, ${\bf e}_\o^i$ and ${\bf e}_{\bar \o}^i$ and hence we arrive at a six-dimensional representation  of $G^{D_4^{\oplus 6}}$.  The corresponding 6-dimensional cycle shape is denoted by $\bar{\Pi}_g$ in Table \ref{tab:chars:eul:6+3}, and the corresponding character denoted by $\bar \chi$. One has $\bar \chi=\chi_1+\chi_3$ in terms of the irreducible representations (cf. Table \ref{tab:chars:irr:6+3}). Alternatively, one might think of the 6-dimensional representation as spanned by the 6 vectors of the form ${\bf e}_1^i+{\bf e}_\o^i+{\bf e}_{\bar \o}^i$. 
Similarly, we also define another character $\chi$ by $\chi_g = \bar \chi_g \epsilon_g$. 
One has $\chi=\chi_2+\chi_4$ in terms of the irreducible representations. Finally, we have the 12-dimensional representation with basis ${\bf e}_1^i-{\bf e}_{\o}^i$ and ${\bf e}_1^i-{\bf e}_{\bar\o}^i$ for $i=1,\dots,6$. We denote the corresponding character by $\check \chi$, given by $\check \chi=\chi_{14}$ in terms of the irreducible characters. 

One can translate the above description of the group action on the hexacode into an action on the root systems $D_4^{\oplus 6}$ in a straightforward way. First one identifies each copy of $\FF_4$ with a copy of $D_4$ and identifies ${\bf e}_0$ with the central node of the dynkin diagram and ${\bf e}_1$, ${\bf e}_\o$ and ${\bf e}_{\bar \o}$ with the three nodes connected to the central node.

\subsection{The Singularities}
\label{sec:Singularities}

As reviewed in \S\ref{subsec:umbral_twining_genera} 
there are 23 different natural ways to decompose the K3 elliptic genus (and twinings thereof) into two parts, corresponding to the 23 Niemeier root systems $X$. The first part is given by the elliptic genus  of the CFTs that describes the singularities associated with $X$.
The second part is the contribution from the umbral moonshine vector-valued mock modular forms $H^X$.
As the umbral group $G^X$ naturally acts on the singularities $X$ as well as the umbral moonshine module, we can generalise the construction and define a $g$-twined weak Jacobi form $\phi^X_g$ as in \eqref{eq:um_twining}.

In this subsection we describe the construction of the twined singularity elliptic genus $\textbf{EG}_g(\t,\z;X)$ for $X=D_4^6$ explicitly, for all $g\in G^{D4^{\oplus 6}} \cong 3.S_6$
This is expressed via \eqref{def:EG_ADE} in terms of the elliptic genus of the $D_4$ supersymmetric minimal model, given by 
\begin{equation}
\label{eq:Z_min}
Z_{\text{minimal}}^{D_4}(\tau,\zeta)=\frac{1}{2}\tr\left(\Omega^{D_4}\cdot\tilde{\chi}(\tau,\zeta)\right)=\frac{\theta_1^2\left(\tau,\frac{2}{3}\z\right)}{\theta_1^2\left(\tau,\frac{1}{3}\z\right)}, 
\end{equation}
where the Cappelli--Itzykson--Zuber \cite{Cappelli:1987xt} omega matrix $\Omega^{D_4}$ is given by 
\begin{equation}
\label{eq:cappelli_twisted}
\left(\Omega^{D_4}\right)_{r,r^\prime}=\begin{cases}
 2\delta_{r,r^\prime(12)}~~~~~~~~~~~~~~~~~~, & r=0,3\mod6 \\ \delta_{r,r^\prime(12)}+\delta_{r,-r^\prime(6)}\delta_{r,r^\prime(4)}, & r=1,5\mod6~~, \\ 
\delta_{r,r^\prime(12)}+\delta_{r,-r^\prime(6)}\delta_{r,r^\prime(4)}, & r=2,4\mod6
\end{cases}
\end{equation}
and $r,r^\prime$ are taken to be in $\ZZ/12$.
Using the property $\tilde{\chi}^r_s(\tau,\zeta)=-\tilde{\chi}^{-r}_s(\tau,\zeta)$ of the parafermion characters, we can rewrite (\ref{eq:Z_min}) as
\be
Z_{\text{minimal}}^{D_4}= \tr_* \left(\widehat \Omega^{D_4} \cdot \tilde{\chi}\right) ,
\ee
where $(\widehat \Omega^{D_4})_{r,s} =( \Omega^{D_4})_{r,s} -( \Omega^{D_4})_{r,-s} $ and is expicitly given by 
\begin{equation}
\widehat{\Omega}^{D_4}=\left(\begin{array}{ccccc} 1 & 0 & 0 & 0 & 1 \\ 0 & 0 & 0 & 0 & 0  \\ 0 & 0 & 2 & 0 & 0 \\ 0 & 0 & 0 & 0 & 0 \\  1 & 0 & 0 & 0 & 1\end{array} \right)
\end{equation}
and we have used the notation  $\tr_*$ to denote tracing over the indices $\lbrace1,2,3,4,5\rbrace$.
Then from (\ref{def:EG_ADE}) this gives the corresponding singularity elliptic genus
\begin{equation}
\label{eq:z_id_final}
\mathbf{EG}(\tau,\zeta;K3)=\tr_* \left(\widehat{\Omega}^{D_4}\cdot\Xi(\tau,\zeta)\right)\end{equation}
where we  have defined
\begin{equation}
\label{eq:xi}
\Xi^r_s(\tau,\zeta):= 
\frac{1}{6} \sum_{a,b \in \ZZ/6\ZZ} q^{a^2} y^{2a} \, \tilde \chi^r_s(\tau,\zeta+a\tau+b)  Z_{L_m}(\tau,\zeta+a\t+b)
\end{equation}
which has integer coefficients in the $q,y$ expansions.

Recall that the automorphism group of the $D_4$ root system is generated by an order 2 element $g_2$ and an order 3 element $g_3$. 
The corresponding action on the minimal model is then captured by $Z_{\text{minimal},g_{2,3}}^{D_4}= \tr_* \left(\widehat \Omega^{D_4}_{g_{2,3}} \cdot \tilde{\chi}\right)$, where the so-called twined Omega matrices for $D_4$ are given by 
\begin{equation}
\widehat{\Omega}^{D_4}_{g_2}=\left(\begin{array}{ccccc} 1 & 0 & 0 & 0 & -1 \\ 0 & 0 & 0 & 0 & 0  \\ 0 & 0 & 0 & 0 & 0 \\ 0 & 0 & 0 & 0 & 0 \\  -1 & 0 & 0 & 0 & 1\end{array} \right)~~,~~~
\widehat{\Omega}^{D_4}_{g_3}=\left(\begin{array}{ccccc} 1 & 0 & 0 & 0 & 1 \\ 0 & 0 & 0 & 0 & 0  \\ 0 & 0 & -1 & 0 & 0 \\ 0 & 0 & 0 & 0 & 0 \\  1 & 0 & 0 & 0 & 1\end{array} \right).
\end{equation}
From this and from the explicit description of the group action of $G^{D_4^{\oplus 6}}$ on the root system $D_4^{\oplus 6}$ given in 
\S\ref{subsec:thegroup}, we conclude that the corresponding twined partition function for 6 copies of $D_4$ singularity is given by
\be\label{eqn:twined_singularity_piece}
{\bf EG}_g(\tau,\zeta;D_4^{\oplus 6}) = \tr_* \left(\widehat \Omega^{D_4^{\oplus 6}}_g \cdot \Xi\right)
\ee
where the $g$-twined omega matrix $\Omega^{D_4^{\oplus 6}}_g$ is given by the group charaters $\chi$, $\bar \chi$, $\check \chi$ discussed in \S\ref{subsec:thegroup} as
\begin{equation}
\widehat{\Omega}^{D_4^{\oplus 6}}_g=\left(\begin{array}{ccccc} \bar{\chi}_g & 0 & 0 & 0 & \chi_g \\ 0 & 0 & 0 & 0 & 0  \\ 0 & 0 & \check{\chi}_g & 0 & 0 \\ 0 & 0 & 0 & 0 & 0 \\  \chi_g & 0 & 0 & 0 & \bar{\chi}_g \end{array} \right)
\end{equation}
for $g\in 3.S_6$. 
The above expression (\ref{eqn:twined_singularity_piece}) makes manifest the $G^{D_4^{\oplus 6}}$-supermodule underlying the singularity CFT.

\subsection{The Mock Modular Forms}
\label{subsec:mock_forms}

It was shown in appendix D of \cite{k3_lattices} that, for any embedding $\iota :G^{D_4^{\oplus6}} \to \Co_0$ we have
\begin{equation}
\label{eq:embedding}
\phi_{\epsilon_g,\i(g)} = \phi^{D_4^{\oplus6}}_{g},
\end{equation}
where we are using definition \eqref{eq:phi_epsilon} with $\e$ given by $\e_g:G^{D_4^{\oplus6}}\rightarrow\lbrace1,-1\rbrace$, the character defined in \S\ref{subsec:thegroup}. The results of \S\ref{sec:The Chiral CFT} and \S\ref{sec:Singularities} amount to a construction of an infinite-dimensional, $\ZZ\times \ZZ$-graded super-module $$W =\bigoplus_{n,\ell \in\ZZ}W_{n,\ell}, ~~W_{n,\ell} = W_{n,\ell}^{(+)}\oplus W_{n,\ell}^{(-)}$$ for $G^{D_4^{\oplus6}}$, such that  its graded super character, defined for a super-module $V=V^{(+)}\oplus V^{(-)}$ as ${\rm Str}_V(g) :={\rm Tr}_{V^{(+)}}(g)-{\rm Tr}_{V^{(-)}}(g) $, gives
\be
 \tilde \phi_{\epsilon_g,g}(\tau,\zeta) = \sum_{n,\ell \in\ZZ} q^n y^\ell \,{\rm Str}_{W_{n,\ell}}(g)     ,
\ee
where
\be
 \tilde \phi_{\epsilon_g,g}(\tau,\zeta) := \phi_{\epsilon_g,g}(\tau,\zeta) - {\bf EG}_g(\tau,\zeta;D_4^{\oplus6})  .
\ee
Attaching the Hilbert space ${\cal H}_{\rm aux}$ of a pair of (auxilliary) periodic bosonic ghosts, on which the group $G^{D_4^{\oplus 6}}$ acts trivially, 
 we obtain
\be
K_{n,\ell} = K_{n,\ell}^{(+)}\oplus K_{n,\ell}^{(-)}, ~~{\text{ with  }}K_{n,\ell}^{(\pm)} = {\cal H}_{\rm aux}\otimes W_{n,\ell}^{(\pm)}. 
\ee
From the discussions above we conclude that its graded characters are given by (cf. \eq{eq:chi_B_R_z}) 
\be
 \sum_{n,\ell \in\ZZ} q^n y^\ell {\rm Str}_{K_{n,\ell}}(g) =   h_{g}(\tau) 
\ee
where 
\be
h_{g}(\tau) := -{\eta^2(\tau)\over \theta_1^2(\tau,\zeta)} \tilde \phi_{\epsilon_g,g}(\tau,\zeta)
\ee
Note that, from \eqref{eq:embedding} and \eqref{eq:um_twining} we see that the function $h_{g}$ is indeed independent of the $\z$-variable. This implies that 
$
{\rm Str}_{K_{n,\ell}}(g) =0 
$
for all $g\in G^{D_4^{\oplus6}}$ unless $\ell=0$, although this is not obvious from the way the virtual representation $K_{n,\ell}$ is constructed. 

Now we shall see how the umbral moonshine mock modular forms $H^{D_4^{\oplus6}}_{g}=(H_{g,r})$, $r=1,3,5$, are recovered from the $G^{D_4^{\oplus6}}$-supermodule $K=\bigoplus_{n,\ell \in\ZZ}K_{n,\ell}$. 
  Let ${z}$ be a non-trivial element in the center subgroup of $G^{D_4^{\oplus6}}$, $\langle z\rangle\cong\ZZ_3$. Then the umbral moonshine mock modular forms are given by
\begin{align} \label{main_def_mod1} 
-{{\theta_{6,3}^1(\t) } \over \eta^4(\t)}H_{g,3}(\t) &=\tfrac{1}{3} (2h_g - h_{zg}- h_{z^2g})(\t) \\\label{main_def_mod2} 
 -{(\theta^1_{6,1}+\epsilon_g  \theta^1_{6,5})(\t)\over \eta^4(\t) }H_{g,1}(\t)& =\tfrac{1}{3} (h_g +h_{zg}+h_{z^2g})(\t)\\ \label{main_def_mod3} 
 -\epsilon_g{(\theta^1_{6,1}+\epsilon_g  \theta^1_{6,5})(\t) \over \eta^4(\t) }H_{g,5}(\t)& = \tfrac{1}{3} (h_g +h_{zg}+h_{z^2g})(\t)
\end{align}
where the unary theta functions $ \theta^1_{6,r}(\t)$ are defined in \eqref{theta_m1}.

Note that despite the apparent factor of $1\over 3$,  the above expressions give 
a construction of $G^{D_4^{\oplus6}}$-supermodules whose graded supertraces coincide with the mock modular forms $H_{g,r}$ for $r=1,3,5$. This is because the right-hand side of (\ref{main_def_mod1}) (resp.  (\ref{main_def_mod2}-\ref{main_def_mod3})) has the interpretation of projecting out the  $G^{D_4^{\oplus6}}$--representations that  factor through $S_6$ (resp.  are faithful representations). 
One can see this explicitly by looking at the character table  (Table \ref{tab:chars:irr:6+3}).
Explicitly, let us define the projection operator $P$ acting acting on any virtual representation $V = \sum_{i=1}^{16} n_i V_i$ of $G^{D_4^{\oplus6}}$, where $n_i\in \ZZ$ and $V_i$ denotes the irreducible representation corresponding to the character $\chi_i$ in  Table \ref{tab:chars:irr:6+3}, by $V\lvert P = \sum_{i=1}^{11} n_i V_i$, and similarly $P' := {\rm{id}} -P$ acting as  $V\lvert P' = \sum_{i=12}^{16} n_i V_i$. Then we have\footnote{Departing from the CFT structure, from the explicit vector space interpretation of the prefactors in \eq{eqn:Hfrommodule} 
one can define infinite-dimensional virtual (trivial) representations $\tilde K_3 = \oplus_{n} \tilde K_{3,n}$ such that $H_{g,3}(\t)  =  \sum_{n, n', \ell \in\ZZ} q^{n+n'} y^\ell \, {\rm Str}_{\tilde K_{3,n}\otimes(K_{n',\ell}\lvert P)}(g) $. Similarly, one can define graded virtual representations
$\tilde K_1$ and $\tilde K_5$ such that $H_{g,r}(\t)  =  \sum_{n, n', \ell \in\ZZ} q^{n+n'} y^\ell \, {\rm Str}_{\tilde K_{r,n}\otimes(K_{n',\ell}\lvert P')}(g) $ for $r=1,5$ .
} 
\begin{align}\label{eqn:Hfrommodule}
-{{\theta_{6,3}^1(\t)  } \over \eta^4(\t)}H_{g,3}(\t) &= \sum_{n,\ell \in\ZZ} q^n y^\ell \, {\rm Str}_{K_{n,\ell}\lvert P}(g)   \\ \label{main_def_module}  
 -{(\theta^1_{6,1}+\epsilon_g  \theta^1_{6,5})(\t) \over \eta^4(\t)} H_{g,1}(\t)&=-\epsilon_g{(\theta^1_{6,1}+\epsilon_g  \theta^1_{6,5})\over \eta^4(\t)} H_{g,5}(\t)=\sum_{n,\ell \in\ZZ} q^n y^\ell \, {\rm Str}_{K_{n,\ell}\lvert P'}(g) .
\end{align}
Note that this construction makes manifest the property described in Conjecture 6.2 in \cite{UMNL}.
The first few coefficients of the mock modular forms $H^{D_4^{\oplus6}}_{g}$ and the corresponding $G^{D_4^{\oplus6}}_{g}$--representations can be found in Appendix C.6  and D.6 of \cite{UMNL}.

\section{Discussion}
\label{sec:discussions}

In this paper we construct a module for the $D_4^{\oplus 6}$ case of umbral moonshine. This is the first time that the module is constructed for a case of umbral moonshine with a sizeable umbral finite group (with $|G^{D_4^{\oplus 6}}|\sim 10^3$, this group is larger than the cases of umbral moonshine where modules have been constructed previously in \cite{Duncan:2014tya,Duncan:2017bhh,Dtypemodule}, where the groups have order dividing 24). 
This is also the first construction of the umbral module which utilises the connection to symmetries of K3 string theory. 
At the same time, there are clearly important open questions remaining. In the following we discuss a few of them. 

\begin{itemize}[leftmargin=*]
\item 
Note that our construction naturally leads to a super module for $G^{D_4^{\oplus 6}}$. However, apart from the virtual representation corresponding to the leading polar term 
(cf. \eq{eqn:umbralmodule}), the umbral module is known to constitute of the even part. It would be nice to make this positivity manifest. 
\item What is the physical or geometric meaning of the chiral CFT ${\cal T}$? 
The relation between the Conway CFT, which is closely related to  ${\cal T}$,  and a specific K3 sigma model has been elucidated  in \cite{derived, reflected, self-dual}. It would be interesting to understand the physical role played by the ghost systems. 
\item An obvious question is whether one can employ a similar construction for the other cases of umbral moonshine.  Note that the chiral CFT ${\cal T}$ has $\text{Spin(24)}$ symmetry which preserves the fermionic and bosonic $U(1)$ ghost currents. It is hence possible to define the regularised twined partition function $\lim_{\eta\to 0^+}Z_g^{\eta}(\t,\zeta_1,\zeta_2)$ (cf. \S\ref{subsec:twinings}) for any element of any of the 23 umbral groups. To make contact with weak Jacobi forms of the type of K3 elliptic genus, one has to specialise the fugacity to $\zeta_1= {1\over 2}$. However, this leads to a finite answer only when taking $\eta\to 0^+$ if $g$ is 4-plane preserving. To construct umbral moonshine modules for cases where not all group elements are 4-plane preserving ($X= A_1^{\oplus24}$, $A_2^{\oplus12}$, and $A_4^{\oplus 6}$), one needs a construction that works with the two-elliptic-variable functions $Z_g(\t,\zeta_1,\zeta_2)$ directly. 
\item Note that the contribution of the vector-valued umbral moonshine mock modular forms $(H^X_{g,r})$ to the twined partition function of the theory ${\cal T}$ is basically given by a single $q$-series ${1\over 2\pi i}{\partial\over \partial \omega} \Psi^X_e(\tau,w)$. See (\ref{eq:um_twining}). 
What allows us to recover from it the individual components $H^X_{g,r}$ of the mock modular forms is the following two facts. First, there are just two independent components in the case $X=D_4^{\oplus 6}$, which can be taken to be $H^X_{g,1}$ and $H^X_{g,3}$. 
Second, the representations underlying the 1st resp. 3rd component have the feature that they factor through $S_6$ resp. are faithful representations. As a result, it is possible to use the projection operator to isolate the contributions from the two independent components from the twined partition function of ${\cal T}$. A similar projection property also holds for other 14 cases of umbral moonshine (cf. Conjecture 6.3 in \cite{UMNL}). 

In view of this, another challenge when attempting to generalise the current construction to other cases of umbral moonshine is how to disentangle the contributions from different components of the vector-valued umbral moonshine mock modular forms $(H^X_{g,r})$ in the twined partition functions for the cases of $X$ with many independent components. 
Recall that an important feature of umbral moonshine is the ``multiplicative relations'' relating  $H^{X'}_{g'}$ and $H^X_g$ for specific pairs of Niemeier root systems $(X, X')$ and group elements $g\in G^X$ and $g'\in G^{X'}$ (cf. \S5.3 of \cite{UMNL}). As we will see in more detail below, these relations together with the projection property enable us to disentangle different components in the vector-valued functions $H^X_g$ in various cases.

\end{itemize}

Finally, we point out that for the cases that $g$ is a 4-plane preserving group element of a umbral group $G^X$, 
many mock modular forms $H^X_g$ for many different $X$ and $g$ can be obtained in a similar way as discussed in \S\ref{sec:module}. 

\vspace{7pt}

\noindent
\underline{\it $A_1^{\oplus 24}$ :} \\
Exactly the same procedure as discussed in the main part of this paper can be used to obtain a supermodule for the group $M_{22}< G^{A_1^{\oplus 24}}$ that is compatible with the $M_{24}$ moonshine, or equivalently the $X=A_1^{\oplus 24}$ case of umbral moonshine. \newline

\noindent
\underline{\it $A_2^{\oplus 12}$:} \\
An analogous procedure, using a projection operator projecting out  representations factoring through $M_{12}< G^{A_2^{\oplus 12}} \cong 2.M_{12}$, recover from twined parition functions (\ref{eqn:def_twinedpf}) the mock modular forms $H^{A_2^{\oplus 12}}_{g}$  for $g\in  2.M_{12}$ that are not in the conjugacy classes $11AB, 12A, 20AB, 22AB$. (Here and below we use the same naming of the conjugacy classes as in \cite{UMNL}.) As a result, one can construct modules for $\tilde G<2.M_{12}$ compatible with the corresponding case of umbral moonshine, for three of the maximal subgroups of $2.M_{12}$. For completeness we list the explicit generators of these three maximal subgroups in terms of permutation groups on 24 objects: 
\begin{equation}\notag
\begin{split}
&G_1=\langle(1,18,5,9,24,16)(2,6,8,11,17,20)(3,12,23,13,22,14)(4,10,19,15,21,7), \\
&~~~~~~~~~~~(1,9)(2,19)(3,13(4,10)(5,24)(6,17)(7,11)(8,23)(12,22)(14,20)(15,21)(16,18)\rangle \\
&G_2=\langle(1,13)(2,19)(3,9)(4,18)(5,21)(6,17)(7,11)(8,20)(10,16)(12,22)(14,23)(15,24), \\
&~~~~~~~~~~~(1,5,11)(2,9,16)(3,18,8)(4,17,7)(6,19,15)(10,12,23)(13,24,20)(14,21,22), \\
&~~~~~~~~~~~(1,9)(2,11)(3,13)(4,15)(5,16)(6,17)(7,19)(8,20)(10,21)(12,22)(14,23)(18,24)\rangle \\
&G_3=\langle(1,12,18)(3,15,6,21,16,14)(4,17,10,5,23,13)(7,20)(8,19)(9,22,24), \\
&~~~~~~~~~~~(1,11,22,24)(2,12,18,9)(3,20,10,19)(4,5)(6,23)(7,13,8,21)(14,17)(15,16)\rangle \\
\end{split}
\end{equation}

\noindent
\underline{\it $A_3^{\oplus 8}$, $A_8^{\oplus 3}$, $E_6^{\oplus 4}$:} \\
Using similar analysis as above, one can recover $H^X_g$ for all elements of $g \in \tilde G<G^X$, for $X= A_3^{\oplus 8}$ and $A_8^{\oplus 3}$. 
In particular, in the $X= A_3^{\oplus 8}$ case we also make use of the multiplicative relations between $H^X_g$ and $H^{X'}_{g'}$, where $X'= A_1^{\oplus 24}$ and $g'\in G^{X'}$, and thereby obtain all $H^X_g$ except for $g\in [8A]$.  In the $X= A_8^{\oplus 3}$ we make use of the multiplicative relations between $H^X_g$ and $H^{X'}_{g'}$, where $X'= A_2^{\oplus 12}$ and $g'\in G^{X'}$, and thereby obtain all $H^X_g$ except for $g\in [3A]$ and $g\in [6A]$. In the $X= E_6^{\oplus 4}$ we make use of the multiplicative relations between $H^X_g$ and $H^{X'}_{g'}$, where $X'= A_2^{\oplus 12}$ and $g'\in G^{X'}$, and thereby obtain all $H^X_g$ except for $g\in [8A]$ and $g\in [8B]$.\\

\noindent
\underline{\it $A_5^{\oplus 4}D_4$, $A_7^{\oplus 2}D_5^{\oplus 2}$ , $A_{11}D_7 E_6$, $A_{17}E_7$, $D_{10}E_7^{\oplus 2}$:} \\
For $X=A_5^{\oplus 4}D_4$ all non-vanishing components of $H^X_g=(H^X_{g,r})$, $r=1,2,\dots,5$, can be recovered from the twined partition functions, by relating them to the umbral moonshine mock modular forms for $X'=D_4^{\oplus 6}$ that we constructed in the main part of the paper, and for the  $X''=A_2^{\oplus 12}$ case that we described above. Explicitly, we have
\be
H^{X'}_{g',1}(\tau)=H^{X}_{g,1} (\tau)+ H^{X}_{g,5}(\tau)
\ee
for the pairs \be\label{ellis5_pair1} (g',g)= (1A,1A/2A), (2A,2B), (2A, 4A), (3B,3A/6A), (4A,8AB),\ee and
\be
H^{X'}_{g'',1}(\tau)=H^{X}_{g,1} (\tau)- H^{X}_{g,5}(\tau)
\ee
for the pairs $(g'',g)= (2B,1A/2A), (2B,2B), (2C, 4A), (6B,3A/6A), (4B,8AB)$.
For the 3rd component we make use the relation
\be
H^{X}_{g,3}(\tau)={1\over 2}H^{X'}_{g',3}(\tau)
\ee
for the same pairs $(g',g)$ as in (\ref{ellis5_pair1}). The even components satisfy
\be
H^{X}_{g,2r} = - H^{X}_{zg,2r} , ~~ r\in \ZZ/3 
\ee
where $z$ denotes a generator of the center subgroup $\langle z\rangle \cong \ZZ_2 < G^X$. 
This forces the even components of the vector-valued mock modular forms $H^{X}_{g}$ to vanish for elements $g$ in conjugacy classes $2B,4A,8AB$. 
The rest of $H^{X}_{g,2r}$  can be recovered by using the relation to $X'' =A_2^{\oplus 12}$ case of umbral moonshine:  
\be
H^{X}_{g,2}(\tau)-H^{X}_{g,4}(\tau)=H^{X'' }_{g',2}(\tau).
\ee
for the pairs $(g,g')=(1A,2B), (3A,6C)$. Note that the two terms on the left-hand side contribute to different powers of $q$ when regarding the whole function as a $q$-series and the above relation is therefore enough to determine both the 2nd and the 4th components of the mock modular forms $H^{X}_{g}$. 

Using similar analysis as above, one can recover all $H^X_g$ for all $g\in G^X$, for $X=A_7^{\oplus 2}D_5^{\oplus 2}$, $A_{11}D_7 E_6$, $A_{17}E_7$, and  $D_{10}E_7^{\oplus 2}$. It would be nice to construct an explicit group action which reproduces these functions.

\centerline{\bf{Acknowledgements}}
\medskip
We are grateful to J. Duncan helpful discussions and comments on an earlier draft.
M.C. and S.H. acknowledge the kind hospitality of the Aspen Center for Physics, which is supported by NSF grant PHY-1066293, as this was being completed. 
The work of V.A. and M.C. is supported by ERC starting grant H2020 ERC StG \#640159. 
S.H. is supported by a Harvard University Golub Fellowship in the Physical Sciences and DOE grant DE-SC0007870.

\appendix

\section{Special functions}
\label{ap:special_functions}
Here we list here the definitions of the Dedekind eta function and the Jacobi theta functions, as well as other theta functions. 
\begin{equation}
\label{eq:special_functions}
\begin{split}
&\eta(\tau):=q^{1/24}\prod_{n=1}^\infty\left(1-q^n\right) \\
&\theta_1(\tau,\z):=-i\sum_{n\in\left(\mathbb{Z}+\frac{1}{2}\right)}(-1)^{n-\frac{1}{2}}y^nq^{n^2/2}= \\
&~~~~~~~~~~=-iq^{1/8}\left(y^{1/2}-y^{-1/2}\right)\prod_{n=1}^\infty\left(1-q^n\right)\left(1-yq^n\right)\left(1-y^{-1}q^n\right) \\
&\theta_2(\tau,\z):=\sum_{n\in\left(\mathbb{Z}+\frac{1}{2}\right)}y^nq^{n^2/2}=q^{1/8}\left(y^{1/2}+y^{-1/2}\right)\prod_{n=1}^\infty\left(1-q^n\right)\left(1+yq^n\right)\left(1+y^{-1}q^n\right) \\
&\theta_3(\tau,\z):=\sum_{n\in\mathbb{Z}}y^nq^{n^2/2}=\prod_{n=1}^\infty\left(1-q^n\right)\left(1+yq^{n-1/2}\right)\left(1+y^{-1}q^{n-1/2}\right) \\
&\theta_4(\tau,\z):=\sum_{n\in\mathbb{Z}}(-1)^ny^nq^{n^2/2}=\prod_{n=1}^\infty\left(1-q^n\right)\left(1-yq^{n-1/2}\right)\left(1-y^{-1}q^{n-1/2}\right) \\
\end{split}
\end{equation}
where $q:=e^{2\pi i\t}$ and $y:=e^{2\pi i\z}$.
Given an $m\in \ZZ_{>0}$ we define the index $m$ theta function:
\be\label{theta_m}
\theta_{m,r}(\t,\z) := \sum_{k=r \,(2m)} q^{k^2\over 4m} y^{k}
\ee
and 
\be\label{theta_m1}
\theta_{m,r}^1(\t) :=\Big( {1\over 2\pi i}{\partial\over \partial \zeta }\theta_{m,r}(\t,\z) \Big)\Big\lvert_{\zeta=0}.
\ee

\section{More on Ghosts}
\label{ap:ghosts}
In this appendix we discuss the ground states of the ghost systems in both the $\PE$ and $\AP$ sectors as well as the action of the $\ZZ_2$ orbifold on them, as a complement to \S\ref{subsec:ghosts}.

\subsection*{The ghost ground states}
\label{ap:ghost_ground_states}
The first thing to note is that for the ghost systems the ordering prescription generally changes when we go from the complex plane, where we use canonical ordering, to the cylinder, where it is natural to use normal ordering. By expressing the Virasoro zero mode in terms of the normal ordering, a constant $B$ will appear in the following way (see \cite{polc}),
\begin{align}
\label{eq:virasoro_ghost_orderings}
L^\gh_0=\sum_n(-n)\canord{\B_n\C_{-n}}=\sum_n(-n)\normord{\B_n\C_{-n}}+B.
\end{align}
We define a ground state on the cylinder as a state that is annihilated by the normal ordered term above, so it will still have weight $B$. Thus, this state will not generally be the $SL(2,\RR)$-invariant vacuum $|0\rangle$. In order to treat both cases together, denote the ground state(s) in the $\AP$ and $\PE$ sectors of either ghost system by $|\Omega_\AP\rangle$ and $|\Omega_\PE\rangle$ respectively. The constant $B$ depends on the central charge and the sector as
\begin{align}
\begin{split}
B_\AP=-\frac{\k}{8}Q^2~,~~B_\PE=\frac{\k}{8}(1-Q^2).
\end{split}
\end{align}
We can also compute the eigenvalues of the ground states under the Virasoro zero mode on the cylinder $L^\gh_{\cyl,0}=L^\gh_0-c_{\B\C}/24$. We notice that the $Q$-dependence cancels and we get
\begin{align}
\begin{split}
L^\gh_{\cyl,0}|\Omega_\AP\rangle=-\frac{\k}{24}|\Omega_\AP\rangle,~~~L^\gh_{\cyl,0}|\Omega_\PE\rangle=\frac{\k}{12}|\Omega_\PE\rangle,
\end{split}
\end{align}
 for any value of $c_{\B\C}$. Note that all the characters we use in this paper are defined in terms of the canonically ordered operators, rather than the normal-ordered ones.

As we have mentioned in \S\ref{subsec:ghosts}, the ghost systems possess a $U(1)$ current $J=-\canord{\B\C}$. Note that $J$ is not a primary field, and the failure to be so is captured by the quantity $Q$:
\begin{align}
\label{ope_J_T}
L^\gh(z)J(w)\sim\frac{Q}{(z-w)^3}+\frac{J(w)}{(z-w)^2}+\frac{\partial J(w)}{z-w}.
\end{align}
Accordingly, upon going to the cylinder the charge operator is shifted as $J_{\cyl,0}=J_0+\frac{Q}{2}$. This $U(1)$ current can be used to define an infinite family of primary operators \cite{friedan}, which will eventually be related to the cylinder ground states. First we introduce a new (bosonic) field $\f$ of zero weight so that $J(z)=\k\pa\f(z)$, which results in the OPE $\phi(z)\phi(w)\sim\k\ln(z-w)$. We then define a vertex operator by $V_q(z):=\canord{e^{q\phi(z)}}$, which is primary and obeys the OPEs
\begin{align}
L^\gh(z)V_q(w)\sim\left[\frac{\frac{1}{2}\k q(q+Q)}{(z-w)^2}+\frac{\pa}{z-w}\right]V_q(w),~~~ J(z)V_q(w)\sim\frac{q}{z-w}~V_q(w).
\end{align}
Acting on $|0\rangle$, this operator defines the state
\begin{align}
|q\rangle:= V_q(0)|0\rangle,
\end{align}
which has the following weight and $U(1)$ charge
\begin{align}
\label{eq:eigenvalues_qvacua}
\begin{split}
L^\gh_0|q\rangle=\frac{1}{2}\k q(q+Q)|q\rangle~,~~J_0|q\rangle=q|q\rangle.
\end{split}
\end{align}
These states can be regarded as vacuum states (for the Fock space) on the plane. Note that each such $q$-vacuum is annihilated by a different set of modes of the ghost fields, depending on the eigenvalue $q$. In particular, we have
\begin{align}
\label{eq:ghosts_general_highest_weight_conditions}
\begin{split}
&\B_r|q\rangle=0~~\forall~r\geq\k q+1-h, \\
&\C_r|q\rangle=0~~\forall~r\geq-\k q+h.
\end{split}
\end{align}
These vacua belong to the periodic sector on the plane if $q\in\ZZ$, and to the anti-periodic sector if $q\in\left(\ZZ+\frac{1}{2}\right)$. One way to see this is from \eqref{eq:ghosts_general_highest_weight_conditions}, since $r\pm h$ is should always be an integer in the periodic sector on the plane, and half-integer in the antiperiodic sector. Also note that the vertex operators $V_{\pm1/2}$ interpolate between the above two sectors, and hence can be regarded as twist fields.

The space of states on the cylinder is built by acting with the creation operators on the ground states $|\Omega_\text{S}\rangle$, where $\text{S}=\lbrace\PE,\AP\rbrace$ denotes the two sectors. These are equal to some of the $q$-vacua described above. By equating the weight $B_\text{S}$  and  the weight of the  $q$-vacua \eqref{eq:eigenvalues_qvacua}, we find that they have the following eigenvalues under $J_0$:
\begin{align}
\label{eq:plane_charges}
q_{\Omega_\AP}=-\frac{1}{2}Q,~~~q_{\Omega_\PE}=-\frac{1}{2}(Q\mp\k).
\end{align}
Accounting for the shift $J_{\cyl,0}=J_0+\frac{Q}{2}$, the corresponding $\text{U}(1)$ charges on the cylinder are
\begin{align}
\begin{split}
&J_{\cyl,0}|\Omega_\AP\rangle=0, \\
&J_{\cyl,0}|\Omega_\PE\rangle=\pm\frac{\k}{2}|\Omega_\PE\rangle.
\end{split}
\end{align}

We see that in the $\AP$ sector we have a single ground state, denoted $|\Omega_\AP^\FE\rangle$ and $|\Omega_\AP^\BO\rangle$ for the fermionic and bosonic system respectively. In the $\PE$ sector of the fermionic ghosts the zero modes $b_0, c_0$ form a Clifford algebra, which results in two degenerate ground states with opposite $U(1)$ charges, which we denote by $|\Omega_{\PE,\pm}^\FE\rangle$. They obey
\begin{align}
\begin{split}
b_0|\Omega_{\PE,-}^{\FE}\rangle=0,~~~b_0|\Omega_{\PE,+}^{\FE}\rangle=|\Omega_{\PE,-}^{\FE}\rangle,~~~c_0|\Omega_{\PE,+}^{\FE}\rangle=0,~~~c_0|\Omega_{\PE,-}^{\FE}\rangle=|\Omega_{\PE,+}^{\FE}\rangle.\\
\end{split}
\end{align}
In the bosonic case, we have to single out one of the two possible ground states in the $\PE$ sector since they do not belong in the same representation of the $\b, \g$ algebra. In other words, the zero modes $\b_0, \g_0$ do not form an analogue of the Clifford algebra of their fermionic counterparts, so one of them must be an annihilator. We choose to use the ground state with positive charge for the torus characters that will follow, which corresponds to $\g_0$ being a creation and $\b_0$ being an annihilation operator.

\subsection*{The ghost orbifold}
\label{ap:ghost_orbifold}
We now treat the $\ZZ_2$ orbifold for both fermionic and bosonic ghosts, generated by $\x\B=-\B$ and $\x\C=-\C$. In terms of the field $\phi$ introduced earlier, the ghost fields are expressed as
\begin{align}
\label{eq:bosonization}
\begin{split}
&b(z)=\canord{e^{-\f(z)}},~~~~~~~~~~~~c(z)=\canord{e^{\f(z)}}, \\
&\beta(z)=\canord{e^{-{\f(z)}}}\pa\l(z),~~~\g(z)=\canord{e^{\f(z)}}\h(z).
\end{split}
\end{align}
where we introduced two auxiliary fields $\h$, $\l$. These form a free fermionic ghost system by themselves, with $h_{\h\l}=1$ and central charge $c_{\h\l}=-2$. They need to be introduced because the Virasoro operator that is build out of $J$,
\begin{align}
T_J=-\left(\frac{1}{2}\canord{JJ}-\frac{1}{2}Q\pa J\right),
\end{align}
is not enough to describe the bosonic ghosts \cite{friedan}. In particular, there is a ``residual" Virasoro operator $T_{-2}$ that needs to be added, so that the total ghost Virasoro operator is given by $L=T_J+T_{-2}$, where $T_{-2}$ is precisely the Virasoro operator of the fermionic ghosts $\h$, $\l$.

One can implement the action of $\x$ on the ``bosonized" form of the ghost fields \eqref{eq:bosonization} in both cases by setting
\begin{align}
\x\f=\f+(2k+1)\pi i,~~k\in\ZZ,~~~~~~~~\x\h=\h,~~~~~\x\l=\l,
\end{align}
Consequently, for the vacuum state $|q\rangle$ with $q\in\ZZ$ we get
\begin{align}
\label{eq:fermionic_parity}
\x|q\rangle=\x\canord{e^{q\f(0)}}|0\rangle=\left\lbrace \begin{array}{cc} +|q\rangle, & q~\text{even} \\ -|q\rangle, & q~\text{odd} \end{array}  \right.,
\end{align}
while for $q\in\ZZ+\frac{1}{2}$ we have
\begin{align}
\label{eq:fermionic_parity_half}
\x|q\rangle=\x\canord{e^{q\f(0)}}|0\rangle=\left\lbrace \begin{array}{cc} +i(-1)^k |q\rangle, & (q-1/2)~~\text{even} \\ -i(-1)^k |q\rangle, & (q-1/2)~~\text{odd} \end{array}  \right..
\end{align}
For convenience, we will choose $k=0$ without loss of generality.

As we have seen, the ground states on the cylinder correspond to certain states $|q\rangle$ on the plane, with charge $q$ under $J_0$. Specifically, for the ground states $|\Omega_{\AP}^\FE\rangle$ and $|\Omega_{\PE,\pm}^\FE\rangle$ in the corresponding sectors of the fermionic system, we have already seen that
\begin{align}
|\Omega^\FE_\AP\rangle=\left|-\frac{Q}{2}\right\rangle,~~~|\Omega_{\PE,\pm}^\FE\rangle=\left|-\frac{1}{2}(Q\mp1)\right\rangle.
\end{align}
From \eqref{eq:fermionic_parity} and \eqref{eq:fermionic_parity_half} we have that
\begin{align}
\x|\Omega^\FE_\AP\rangle=e^{\pi i\left(h-\frac{1}{2}\right)}|\Omega^\FE_\AP\rangle,~~~~~\x|\Omega^\FE_{\PE,\pm}\rangle=\pm e^{\pi ih}|\Omega^\FE_{\PE,\pm}\rangle.
\end{align}
Combining the $\x$ action on the ground states and the oscillators, we get
\begin{align}
\begin{split}
&\til\chi^\AP_\FE(\tau,\zeta):=\tr_\AP\left[\x~y^{J_{\cyl,0}}q^{L^{\gh}_{\cyl,0}}\right] \\
&~~~~~~~=e^{\pi i\left(h-\frac{1}{2}\right)}q^{-\frac{1}{24}}\prod_{n=1}^\infty\left(1-yq^{n-1/2}\right)\left(1-y^{-1}q^{n-1/2}\right)=(-1)^{h-\frac{1}{2}}~\chi_\FE^\AP(\tau,\zeta+1/2),
\end{split}
\end{align}
\begin{align}
\begin{split}
&\til\chi^\PE_\FE(\tau,\zeta):=\tr_\PE\left[\x~y^{J_{\cyl,0}}q^{L^{\gh}_{\cyl,0}}\right] \\
&~~~~~~~=e^{\pi ih}q^{\frac{1}{12}}\left(y^{1/2}-y^{-1/2}\right)\prod_{n=1}^\infty\left(1-yq^n\right)\left(1-y^{-1}q^n\right)=(-1)^{h-\frac{1}{2}}~\chi_\FE^\PE(\tau,\zeta+1/2),
\end{split}
\end{align}

Similarly, on the bosonic system ground states
\begin{align}
|\Omega^\BO_\AP\rangle=\left|-\frac{Q}{2}\right\rangle,~~~|\Omega^\BO_\PE\rangle=\left|-\frac{1}{2}(Q-1)\right\rangle,
\end{align}
the orbifold acts as
\begin{align}
\x|\Omega^\BO_\AP\rangle=e^{\pi i\left(-h+\frac{1}{2}\right)}|\Omega^\BO_\AP\rangle,~~~~~\x|\Omega^\BO_{\PE}\rangle=e^{\pi i\left(-h+1\right)}|\Omega^\BO_{\PE}\rangle.
\end{align}
Thus we calculate
\begin{align}
\begin{split}
&\til\chi^\AP_\BO(\tau,\zeta):=\tr_\AP\left[\x~y^{J_{\cyl,0}}q^{L^{\gh}_{\cyl,0}}\right] \\
&~~~=e^{\pi i\left(-h+\frac{1}{2}\right)}q^{1/24}\prod_{n=1}^\infty\left(1+yq^{n-1/2}\right)^{-1}\left(1+y^{-1}q^{n-1/2}\right)^{-1}=(-1)^{3h+\frac{1}{2}}~\chi_\BO^\AP(\tau,\zeta+1/2),
\end{split}
\end{align}
\begin{align}
\begin{split}
&\til\chi^\PE_\BO(\tau,\zeta):=\tr_\PE\left[\x~y^{J_{\cyl,0}}q^{L^{\gh}_{\cyl,0}}\right] \\
&~~~=e^{\pi i\left(-h+1\right)}q^{-1/12}y^{1/2}(1+y)^{-1}\prod_{n=1}^\infty\left(1+yq^n\right)^{-1}\left(1+y^{-1}q^n\right)^{-1}=(-1)^{3h+\frac{1}{2}}~\chi_\BO^\PE(\tau,\zeta+1/2).
\end{split}
\end{align}

Also note that all traces over the ghost Hilbert space involve defining a dual Fock space: corresponding to each in-state $|x\rangle=\prod_{i}\C_{-r_i}\prod_j\B_{-s_j}|q\rangle$ there us an out-state $\langle y|=\langle q^\prime|\left(\prod_{i}\C_{-r_i}\prod_j\B_{-s_j}\right)^t$ such that their inner product is $\langle y|x\rangle=1$. Due to the charge asymmetry \cite{friedan}, namely $J_0^t=-J_0-Q$, the dual to the vertex state $|q\rangle$ is $\langle-q-Q|:=\langle0|\canord{e^{(-q-Q)\phi(z)}}$, while the duals of the oscillator modes are $\C_{-r}^t=\B_r$ and $\B_{-r}^t=\C_r$. The latter is compatible with transposing the (anti-) commutation relations $\lbrace\B_{r},\C_{-r}\rbrace_\kappa=\kappa$, and we have have $L_0^t=L_0$.

\clearpage

\section{Character Tables}

\begin{table}[h!]
\begin{center}
\caption{\label{tbl:characters}Character table of ${G}^{\rs}\simeq 3.\Sym_6$, $\rs=D_4^6$}\label{tab:chars:irr:6+3}
\smallskip
\begin{small}
\begin{tabular}{c|c|rrrrrrrrrrrrrrrrrrrrrrrrrrr}\toprule
$[g]$&FS&   1A&   3A&   2A&   6A&   3B&   3C&   4A&   12A&   5A&   15A&   15B&   2B&   2C&   4B&	6B&	6C\\
	\midrule
$[g^2]$&&	1A&	3A&	1A&3A&	3B&3C&2A&6A&5A&15A&15B&1A&1A&2A&3B&3C\\
$[g^3]$&&	1A&	1A&	2A&2A&	1A&1A&4A&4A&5A&5A&5A&2B&2C&4B&2B&2C\\
$[g^5]$&&	1A&	3A&	2A&	6A&3B&3C&4A&12A&1A&3A&3A&2B&2C&4B&6B&6C\\
	\midrule
${\chi}_{1}$&$+$&   $1$&   $1$&   $1$&   $1$&   $1$&   $1$&   $1$&   $1$&   $1$&   $1$&   $1$&   $1$&   $1$&   $1$&   $1$&   $1$\\
${\chi}_{2}$&$+$&   $1$&   $1$&   $1$&   $1$&   $1$&   $1$&   $1$&   $1$&   $1$&   $1$&   $1$&   $-1$&   $-1$&   $-1$&   $-1$&   $-1$\\
${\chi}_{3}$&$+$&   $5$&   $5$&   $1$&   $1$&   $2$&   $-1$&   $-1$&   $-1$&   $0$&   $0$&   $0$&   $3$&   $-1$&   $1$&   $0$&   $-1$\\
${\chi}_{4}$&$+$&   $5$&   $5$&   $1$&   $1$&   $2$&   $-1$&   $-1$&   $-1$&   $0$&   $0$&   $0$&   $-3$&   $1$&   $-1$&   $0$&   $1$\\
${\chi}_{5}$&$+$&   $5$&   $5$&   $1$&   $1$&   $-1$&   $2$&   $-1$&   $-1$&   $0$&   $0$&   $0$&   $-1$&   $3$&   $1$&   $-1$&   $0$\\
${\chi}_{6}$&$+$&   $5$&   $5$&   $1$&   $1$&   $-1$&   $2$&   $-1$&   $-1$&   $0$&   $0$&   $0$&   $1$&   $-3$&   $-1$&   $1$&   $0$\\
${\chi}_{7}$&$+$&   $16$&   $16$&   $0$&   $0$&   $-2$&   $-2$&   $0$&   $0$&   $1$&   $1$&   $1$&   $0$&   $0$&   $0$&   $0$&   $0$\\
${\chi}_{8}$&$+$&   $9$&   $9$&   $1$&   $1$&   $0$&   $0$&   $1$&   $1$&   $-1$&   $-1$&   $-1$&   $3$&   $3$&   $-1$&   $0$&   $0$\\
${\chi}_{9}$&$+$&   $9$&   $9$&   $1$&   $1$&   $0$&   $0$&   $1$&   $1$&   $-1$&   $-1$&   $-1$&   $-3$&   $-3$&   $1$&   $0$&   $0$\\
${\chi}_{10}$&$+$&   $10$&   $10$&   $-2$&   $-2$&   $1$&   $1$&   $0$&   $0$&   $0$&   $0$&   $0$&   $2$&   $-2$&   $0$&   $-1$&   $1$\\
${\chi}_{11}$&$+$&   $10$&   $10$&   $-2$&   $-2$&   $1$&   $1$&   $0$&   $0$&   $0$&   $0$&   $0$&   $-2$&   $2$&   $0$&   $1$&   $-1$\\
${\chi}_{12}$&$\circ$&   $6$&   $-3$&   $-2$&   $1$&   $0$&   $0$&   $2$&   $-1$&   $1$&   $b_{15}$&   $\overline{b_{15}}$&   $0$&   $0$&   $0$&   $0$&   $0$\\
${\chi}_{13}$&$\circ$&   $6$&   $-3$&   $-2$&   $1$&   $0$&   $0$&   $2$&   $-1$&   $1$&   $\overline{b_{15}}$&   $b_{15}$&   $0$&   $0$&   $0$&   $0$&   $0$\\
${\chi}_{14}$&$+$&   $12$&   $-6$&   $4$&   $-2$&   $0$&   $0$&   $0$&   $0$&   $2$&   $-1$&   $-1$&   $0$&   $0$&   $0$&   $0$&   $0$\\
${\chi}_{15}$&$+$&   $18$&   $-9$&   $2$&   $-1$&   $0$&   $0$&   $2$&   $-1$&   $-2$&   $1$&   $1$&   $0$&   $0$&   $0$&   $0$&   $0$\\
${\chi}_{16}$&$+$&   $30$&   $-15$&   $-2$&   $1$&   $0$&   $0$&   $-2$&   $1$&   $0$&   $0$&   $0$&   $0$&   $0$&   $0$&   $0$&   $0$\\\bottomrule
\end{tabular}
\end{small}
\end{center}
\end{table}

\begin{table}
\vspace{-6cm}
\begin{center}
\caption{Twisted Euler characters and Frame shapes at $\ll=6+3$, $\rs=D_4^6$}\label{tab:chars:eul:6+3}
\begin{tabular}{l@{\;}|@{\;}r@{\;}r@{\;}r@{\;}r@{\;}r@{\;}r@{\;}r@{\;}r@{\;}r@{\;}r@{\;}r@{\;}r@{\;}r@{\;}r@{\;}r}\toprule
$[g]$&   		1A&		3A&   	2A&   	6A&			3B&			3C&   	4A&	12A&	5A&	15AB&	2B&	2C&	4B&	6B&	6C\\ 
	\midrule
$n_g|h_g$&		$1|1$&	$1|3$&	$2|1$&	$2|3$&		$3|1$&		$3|3$&	$4|2$&	$4|6$&	$5|1$&	$5|3$&	$2|1$&	$2|2$&	$4|1$&	$6|1$&	$6|6$\\	
	\midrule
$\bar{\chi}_g$&
			$6$&	$6$&	$2$&	$2$&	$3$&	$0$&	$0$&	$0$&	$1$&	$1$&	$4$&	$0$&	$2$&	$1$&	$0$\\
${\chi}_g$&
			$6$&	$6$&	$2$&	$2$&	$3$&	$0$&	$0$&	$0$&	$1$&	$1$&	$-4$&	$0$&	$-2$&	$-1$&	$0$\\
$\check{\chi}_g$&
			$12$&	$-6$&	$4$&	$-2$&	$0$&	$0$&	$0$&	$0$&	$2$&	$-1$&	$0$&	$0$&	$0$&	$0$&	$0$\\
			\midrule
$\bar{\Pi}_g$&
			$1^6$&	$1^6$&	$1^22^2$&	$1^22^2$&	$1^33^1$&	$3^2$&	$2^14^1$&	$2^14^1$&	$1^15^1$&	$1^15^1$&	$1^42^1$&	$2^3$&$1^24^1$&$1^12^13^1$&	$6^1$\\
$\widetilde{\Pi}_g$&
			$1^{24}$&	$1^63^6$&	$1^82^8$&	$1^22^23^26^2$&	$1^63^6$&	$3^8$&	$2^44^4$&$2^14^16^112^1$&	$1^45^4$&$1^13^15^115^1$&$1^22^8$&$2^{12}$&$1^42^24^4$&$1^22^23^26^2$&$6^4$
	\\\bottomrule
\end{tabular}
\smallskip
\end{center}
\end{table}


\clearpage
  
  \addcontentsline{toc}{section}{References}
  \begin{thebibliography}{9}
  
   \bibitem{CN} 
  J.~H.~Conway and S.~P.~Norton,
  ``Monstrous Moonshine,''
  Bull.\ London Math.\ Soc.\  {\bf 11}, no. 3, 308 (1979).
  doi:10.1112/blms/11.3.308
  
      \bibitem{FLM} 
  I.~Frenkel, J.~Lepowsky and A.~Meurman,
  ``Vertex Operator Algebras And The Monster,''
  BOSTON, USA: ACADEMIC (1988) 508 P. (PURE AND APPLIED MATHEMATICS, 134)
  
  \bibitem{EOT} 
  T.~Eguchi, H.~Ooguri and Y.~Tachikawa,
  ``Notes on the K3 Surface and the Mathieu group $M_{24}$,''
  Exper.\ Math.\  {\bf 20}, 91 (2011)
  doi:10.1080/10586458.2011.544585
  [arXiv:1004.0956 [hep-th]].
  
\bibitem{Dabholkar:2012nd}
A.~Dabholkar, S.~Murthy, and D.~Zagier, ``{Quantum Black Holes, Wall Crossing,
  and Mock Modular Forms},''
\href{http://arxiv.org/abs/1208.4074}{{\tt arXiv:1208.4074 [hep-th]}}.

  
    \bibitem{UM} 
  M.~C.~N.~Cheng, J.~F.~R.~Duncan and J.~A.~Harvey,
  ``Umbral Moonshine,''
  Commun.\ Num.\ Theor.\ Phys.\  {\bf 08}, 101 (2014)
  doi:10.4310/CNTP.2014.v8.n2.a1
  [arXiv:1204.2779 [math.RT]].
  
  \bibitem{UMNL} 
  M.~C.~N.~Cheng, J.~F.~R.~Duncan and J.~A.~Harvey,
  ``Umbral Moonshine and the Niemeier Lattices,''
  arXiv:1307.5793 [math.RT].
  
  
  \bibitem{Dtypemodule}
  J.~F.~R.~Duncan and M.~C.~N.~Cheng, 
  ``Meromorphic Jacobi Forms of Half-Integral Index and Umbral Moonshine Modules,''  arXiv:1707.01336 [math.RT].
  
    \bibitem{thompson}
  J.~A.~Harvey and B.~C.~Rayhaun,
  ``Traces of Singular Moduli and Moonshine for the Thompson Group,''
  Commun.\ Num.\ Theor.\ Phys.\  {\bf 10}, 23 (2016)
  doi:10.4310/CNTP.2016.v10.n1.a2
  [arXiv:1504.08179 [math.RT]].
  
  \bibitem{review} 
  J.~F.~R.~Duncan, M.~J.~Griffin and K.~Ono,
  ``Moonshine,''
  Research in the Mathematical Sciences (2015) 2:11
  [arXiv:1411.6571 [math.RT]].
  
  \bibitem{Gannon:2012ck} 
  T.~Gannon,
  ``Much ado about Mathieu,''
  Adv.\ Math.\  {\bf 301}, 322 (2016)
  doi:10.1016/j.aim.2016.06.014
  [arXiv:1211.5531 [math.RT]].
  
  
  \bibitem{Duncan:2015rga} 
  J.~F.~R.~Duncan, M.~J.~Griffin and K.~Ono,
  ``Proof of the Umbral Moonshine Conjecture,''
  arXiv:1503.01472 [math.RT].
  
  \bibitem{Duncan:2014tya} 
  J.~F.~R.~Duncan and J.~A.~Harvey,
  ``The Umbral Moonshine Module for the Unique Unimodular Niemeier Root System,''
  Alg. Number Th. 11 (2017) 505-535
  doi:10.2140/ant.2017.11.505
  [arXiv:1412.8191 [math.RT]].
  
  \bibitem{Duncan:2017bhh} 
  J.~F.~R.~Duncan and A.~O'Desky,
  ``Super Vertex Algebras, Meromorphic Jacobi Forms and Umbral Moonshine,''
  arXiv:1705.09333 [math.RT].
  
  \bibitem{GHV} 
  M.~R.~Gaberdiel, S.~Hohenegger and R.~Volpato,
  ``Symmetries of K3 sigma models,''
  Commun.\ Num.\ Theor.\ Phys.\  {\bf 6}, 1 (2012)
  doi:10.4310/CNTP.2012.v6.n1.a1
  [arXiv:1106.4315 [hep-th]].
  
  
  
  \bibitem{UMK3} 
  M.~C.~N.~Cheng and S.~Harrison,
  ``Umbral Moonshine and K3 Surfaces,''
  Commun.\ Math.\ Phys.\  {\bf 339}, no. 1, 221 (2015)
  doi:10.1007/s00220-015-2398-5
  [arXiv:1406.0619 [hep-th]].
  
  \bibitem{OV} 
  H.~Ooguri and C.~Vafa,
  ``Two-dimensional black hole and singularities of CY manifolds,''
  Nucl.\ Phys.\ B {\bf 463}, 55 (1996)
  doi:10.1016/0550-3213(96)00008-9
  [hep-th/9511164].
  

  \bibitem{Troost:2010ud} 
  J.~Troost,
  ``The non-compact elliptic genus: mock or modular,''
  JHEP {\bf 1006}, 104 (2010)
  doi:10.1007/JHEP06(2010)104
  [arXiv:1004.3649 [hep-th]].
  
  \bibitem{Eguchi:2010cb} 
  T.~Eguchi and Y.~Sugawara,
  ``Non-holomorphic Modular Forms and SL(2,R)/U(1) Superconformal Field Theory,''
  JHEP {\bf 1103}, 107 (2011)
  doi:10.1007/JHEP03(2011)107
  [arXiv:1012.5721 [hep-th]].
  
  \bibitem{Ashok:2011cy} 
  S.~K.~Ashok and J.~Troost,
  ``A Twisted Non-compact Elliptic Genus,''
  JHEP {\bf 1103}, 067 (2011)
  doi:10.1007/JHEP03(2011)067
  [arXiv:1101.1059 [hep-th]].
  
  \bibitem{Murthy:2013mya} 
  S.~Murthy,
  ``A holomorphic anomaly in the elliptic genus,''
  JHEP {\bf 1406}, 165 (2014)
  doi:10.1007/JHEP06(2014)165
  [arXiv:1311.0918 [hep-th]].
  
  \bibitem{Ashok:2013pya} 
  S.~K.~Ashok, N.~Doroud and J.~Troost,
  ``Localization and real Jacobi forms,''
  JHEP {\bf 1404}, 119 (2014)
  doi:10.1007/JHEP04(2014)119
  [arXiv:1311.1110 [hep-th]].
  
  \bibitem{Harvey:2014nha} 
  J.~A.~Harvey, S.~Lee and S.~Murthy,
  ``Elliptic genera of ALE and ALF manifolds from gauged linear sigma models,''
  JHEP {\bf 1502}, 110 (2015)
  doi:10.1007/JHEP02(2015)110
  [arXiv:1406.6342 [hep-th]].
  
  
  \bibitem{Cappelli:1987xt}
A.~Cappelli, C.~Itzykson, and J.~Zuber, ``{The ADE Classification of Minimal
  and A1(1) Conformal Invariant Theories},''
\href{http://dx.doi.org/10.1007/BF01221394}{{\em Commun.Math.Phys.} {\bf 113}
  (1987)  1}.
  
   \bibitem{witten:LG}
E. Witten, ``{On the Landau-Ginzburg Description of $N=2$ Minimal Models},'' \\
\emph{Int.J.Mod.Phys.A} \textbf{9:4783-4800,1994} (Int.J.Mod.Phys.A9:4783-4800,1994), \\\href{https://arxiv.org/abs/hep-th/9304026}{hep-th/9304026}.
  
  \bibitem{Witten:1991yr}
E.~Witten, ``{On string theory and black holes},''
\href{http://dx.doi.org/10.1103/PhysRevD.44.314}{{\em Phys.Rev.} {\bf D44}
  (1991)  314--324}.
  
  \bibitem{KYY}
T.~Kawai, Y.~Yamada, and S.-K. Yang, ``Elliptic genera and n=2 superconformal
  field theory,'' {\em Nucl.Phys.B} {\bf 414:191-212,1994}
  (Nucl.Phys.B414:191-212,1994)  ,
  \href{http://arxiv.org/abs/hep-th/9306096}{{\tt hep-th/9306096}}.
  
  \bibitem{LG} 
  M.~C.~N.~Cheng, F.~Ferrari, S.~M.~Harrison and N.~M.~Paquette,
  ``Landau-Ginzburg Orbifolds and Symmetries of K3 CFTs,''
  JHEP {\bf 1701}, 046 (2017)
  doi:10.1007/JHEP01(2017)046
  [arXiv:1512.04942 [hep-th]].
  
    
  \bibitem{k3_lattices} 
  M.~C.~N.~Cheng, S.~M.~Harrison, R.~Volpato and M.~Zimet,
  ``K3 String Theory, Lattices and Moonshine,''
  arXiv:1612.04404 [hep-th].
  
  \bibitem{Paquette:2017gmb} 
  N.~M.~Paquette, R.~Volpato and M.~Zimet,
  ``No More Walls! A Tale of Modularity, Symmetry, and Wall Crossing for 1/4 BPS Dyons,''
  JHEP {\bf 1705}, 047 (2017)
  doi:10.1007/JHEP05(2017)047
  [arXiv:1702.05095 [hep-th]].
  
  
  \bibitem{optimal}
  Miranda C. N. Cheng, John F. R. Duncan,
  ``Optimal Mock Jacobi Theta Functions'',
  \href{https://arxiv.org/abs/1605.04480}{arXiv:1605.04480}  
  
  
  \bibitem{weight_one_jacobi}
  Miranda C. N. Cheng, John F. R. Duncan, Jeffrey A. Harvey,
  ``Weight One Jacobi Forms and Umbral Moonshine'',
  \href{https://arxiv.org/abs/1703.03968}{arXiv:1703.03968}
  
  
  
  
  
  
  
  
  
  \bibitem{friedan} 
  D.~Friedan, E.~J.~Martinec and S.~H.~Shenker,
  ``Conformal Invariance, Supersymmetry and String Theory,''
  \href{http://www.physics.rutgers.edu/~friedan/papers/Nucl_Phys_B271_93_1986.pdf}{Nucl.\ Phys.\ B {\bf 271}, 93 (1986).}
  
  \bibitem{Kausch} 
  H.~G.~Kausch,
  ``Curiosities at c = -2,''
  hep-th/9510149.
  
  
  \bibitem{Kausch2} 
  H.~G.~Kausch,
  ``Symplectic fermions,''
  Nucl.\ Phys.\ B {\bf 583}, 513 (2000)
  doi:10.1016/S0550-3213(00)00295-9
  [hep-th/0003029].
  
  \bibitem{ghosts} 
  S.~Guruswamy and A.~W.~W.~Ludwig,
  ``Relating $c < 0$ and $c > 0$ conformal field theories'',
  Nucl.\ Phys.\ B {\bf 519}, 661 (1998)
  [hep-th/9612172].
  
  \bibitem{ghost_vertex}
  W. Eholzer, L. Feher, A. Honecker,
  ``Ghost Systems: A Vertex Algebra Point of View",
   Nucl.Phys. B518 (1998) 669-688 hep-th/9708160 DAMTP-97-85, SISSA-104-97-EP,
  \href{https://arxiv.org/abs/hep-th/9708160}{hep-th/9708160}
  
  \bibitem{ghost_bosonic}
  David Ridout, Simon Wood,
  ``Bosonic Ghosts at $c=2$ as a Logarithmic CFT",
  Lett. Math. Phys. 105 (2015) 279–307,
  \href{https://arxiv.org/abs/1408.4185v3}{arXiv:1408.4185v3}
  
  
  \bibitem{conway} 
  J.~F.~R.~Duncan and S.~Mack-Crane,
  ``The Moonshine Module for Conway?s Group,''
  SIGMA {\bf 3}, e10 (2015)
  doi:10.1017/fms.2015.7
  [arXiv:1409.3829 [math.RT]].
  
  \bibitem{derived} 
  J.~F.~R.~Duncan and S.~Mack-Crane,
  ``Derived Equivalences of K3 Surfaces and Twined Elliptic Genera,''
  arXiv:1506.06198 [math.RT].
  
  \bibitem{m5} 
 M.~C.~N.~Cheng, X.~Dong, J.~F.~R.~Duncan, S.~Harrison, S.~Kachru and T.~Wrase,
  ``Mock Modular Mathieu Moonshine Modules,''
  Research in the Mathematical Sciences (2015) 2:13
  doi:10.1186/s40687-015-0034-9
  [arXiv:1406.5502 [hep-th]].

  
  \bibitem{superconway} 
  J. F. Duncan,
  ``Super-moonshine for Conway’s largest sporadic group,''
  Duke Math. J., 139(2):255-315 (2007)
  [arXiv:math/0502267].
  
  \bibitem{polc} 
  J. Polchinski,
  ``String theory, Vol 1: An introduction of the bosonic string and Vol 2:
Superstring theory and beyond,''
  Cambridge University Press cop. 1998
  
  
\bibitem{GMM}
  M.~R.~Gaberdiel, D.~Persson, H.~Ronellenfitsch and R.~Volpato,
  ``Generalized Mathieu Moonshine,''
  Commun.\ Num.\ Theor Phys.\  {\bf 07} (2013) 145
  doi:10.4310/CNTP.2013.v7.n1.a5
  [arXiv:1211.7074 [hep-th]].
  
\bibitem{GUM}
  M.~C.~N.~Cheng, P.~de Lange and D.~P.~Z.~Whalen,
  ``Generalised Umbral Moonshine,''
  arXiv:1608.07835 [math.RT].
  
  
   \bibitem{wilson} 
   R. A. Wilson,
  ``The Finite Simple Groups'',
   Springer-Verlag London Limited, 2009.
  
  \bibitem{sphere_packing}
  J. H. Conway, N. J. A. Sloane,
  ``Sphere Packings, Lattices and Groups'',
   Springer, 1999.
   
   
   \bibitem{reflected}
  A. Taormina, K. Wendland,
  ``The Conway Moonshine Module is a Reflected K3 Theory'',
   arXiv:1704.03813
   
   \bibitem{self-dual}
  T. Creutzig, J. F. R. Duncan, W. Riedler,
  ``Self-Dual Vertex Operator Superalgebras and Superconformal Field Theory'',
   arXiv:1704.03678
   
   
   

\end{thebibliography}
\end{document}